\documentclass[twocolumn]{aastex701}
\usepackage{comment}
\usepackage{booktabs}
\usepackage{hyperref}

\newcommand{\CII}{C~{\sc ii}}

\newcommand{\SII}{S~{\sc ii}}

\newcommand{\OI}{O~{\sc i}}

\newcommand{\CaII}{Ca~{\sc ii}}

\newcommand{\SiII}{Si~{\sc ii}}
\newcommand{\SiIII}{Si~{\sc iii}}

\newcommand{\TiII}{Ti~{\sc ii}}

\newcommand{\FeII}{Fe~{\sc ii}}
\newcommand{\FeIII}{Fe~{\sc iii}}

\begin{document}

\title{The New Status Qvo? SN 2021qvo is Another 2003fg-like Type Ia Supernova with a Rising Light-Curve Bump}

\correspondingauthor{I.~A.~Abreu Paniagua}
\email{iabreu\_paniagua@uri.edu}

\author[0009-0007-4578-109X]{I.~A.~Abreu~Paniagua}
\affiliation{Institute for Astronomy, University of Hawai‘i, 2680 Woodlawn Drive, Honolulu, HI 96822, USA}
\affiliation{Department of Physics, University of Rhode Island, 2 Lippitt Rd, Kingston, RI, 02881, USA}
\email{iabreu\_paniagua@uri.edu}

\author[0000-0003-3953-9532]{W.~B.~Hoogendam}
\altaffiliation{NSF Graduate Research Fellow}
\affiliation{Institute for Astronomy, University of Hawai‘i, 2680 Woodlawn Drive, Honolulu, HI 96822, USA}
\email{willemh@hawaii.edu}

\author[0000-0002-6230-0151]{D.~O.~Jones}
\affiliation{Institute for Astronomy, University of Hawai‘i, 640 N.\ Aohoku Pl., Hilo, HI 96720, USA}
\email{dojones@hawaii.edu}

\author[0000-0001-9494-179X]{G.~Dimitriadis}
\affiliation{Department of Physics, Lancaster University, Lancaster, LA1 4YB, UK}
\email{g.dimitriadis@lancaster.ac.uk}

\author[0000-0002-2445-5275]{R.~J.~Foley}
\affiliation{Department of Astronomy and Astrophysics, University of California, Santa Cruz, CA 95064, USA}
\email{foley@ucsc.edu}

\author[0000-0002-8526-3963]{C.~Gall}
\affiliation{DARK, Niels Bohr Institute, University of Copenhagen, Jagtvej 128, 2200 Copenhagen, Denmark}
\email{christa.gall@nbi.ku.dk}

\author[0000-0003-3615-9593]{J.~O'Brien}
\affiliation{Department of Astronomy, University of Illinois at Urbana-Champaign, 1002 W. Green St., IL 61801, USA}
\email{jackob@illinois.edu}

\author[0000-0002-5748-4558]{K.~Taggart}
\affiliation{Department of Astronomy and Astrophysics, University of California, Santa Cruz, CA 95064, USA}
\email{k.taggart@ucsc.edu}

\author[0000-0002-4269-7999]{C.~R.~Angus}
\affiliation{Astrophysics Research Centre, School of Mathematics and Physics, Queen’s University Belfast, Belfast BT7 1NN, UK}
\email{c.angus@qub.ac.uk}

\author[0000-0002-5221-7557]{C. Ashall}
\affiliation{Institute for Astronomy, University of Hawai‘i, 2680 Woodlawn Drive, Honolulu, HI 96822, USA}
\email{cashall@hawaii.edu}

\author[0000-0002-4449-9152]{K.~Auchettl}
\affiliation{Department of Astronomy and Astrophysics, University of California, Santa Cruz, CA 95064, USA}
\affiliation{School of Physics, The University of Melbourne, VIC 3010, Australia}
\email{katie.auchettl@unimelb.edu.au}

\author[0000-0003-4263-2228]{D.~A.~Coulter}
\affiliation{Department of Physics and Astronomy, The Johns Hopkins University, Baltimore, MD 21218, USA}
\affiliation{Space Telescope Science Institute, Baltimore, MD 21218, USA}
\email{dcoulter@stsci.edu}

\author[0000-0002-5680-4660]{K.~W.~Davis}
\affiliation{Department of Astronomy and Astrophysics, University of California, Santa Cruz, CA 95064, USA}
\email{kywdavis@ucsc.edu}

\author[0000-0001-5486-2747]{T.~de Boer}
\affiliation{Institute for Astronomy, University of Hawai‘i, 2680 Woodlawn Drive, Honolulu, HI 96822, USA}
\email{tdeboer@hawaii.edu}

\author[0000-0003-3429-7845]{A.~Do}
\affiliation{Institute of Astronomy and Kavli Institute for Cosmology, Madingley Road, Cambridge, CB3 0HA, UK}
\email{ajmd6@cam.ac.uk}

\author[0000-0003-1015-5367]{H.~Gao}
\affiliation{Institute for Astronomy, University of Hawai‘i, 2680 Woodlawn Drive, Honolulu, HI 96822, USA}
\email{hgao@hawaii.edu}

\author[0000-0001-9695-8472]{L.~Izzo}
\affiliation{DARK, Niels Bohr Institute, University of Copenhagen, Jagtvej 128, 2200 Copenhagen, Denmark}
\affiliation{INAF, Osservatorio Astronomico di Capodimonte, Salita Moiariello 16, I-80131 Napoli, Italy}
\email{luca.izzo@nbi.ku.dk}

\author[0000-0002-7272-5129]{C.-C.~Lin}
\affiliation{Institute for Astronomy, University of Hawai‘i, 2680 Woodlawn Drive, Honolulu, HI 96822, USA}
\email{cclin33@hawaii.edu}

\author[0000-0002-9438-3617]{T.~B.~Lowe}
\affiliation{Institute for Astronomy, University of Hawai‘i, 2680 Woodlawn Drive, Honolulu, HI 96822, USA}
\email{tlowe@hawaii.edu}

\author[0000-0001-8483-9089]{Z.~Lai}
\affiliation{Department of Astronomy and Astrophysics, University of California, Santa Cruz, CA 95064, USA}
\affiliation{Las Cumbres Observatory, 6740 Cortona Drive, Suite 102, Goleta, CA 93117-5575, USA}
\email{zlai@sfsu.edu}

\author[0009-0005-1871-7856]{R.~Kaur}
\affiliation{Department of Astronomy and Astrophysics, University of California, Santa Cruz, CA 95064, USA}
\email{ravkaur@ucsc.edu}

\author[0009-0005-5121-2884]{M.~Y.~Kong}
\affiliation{Institute for Astronomy, University of Hawai‘i, 2680 Woodlawn Drive, Honolulu, HI 96822, USA}
\email{ykong2@hawaii.edu}

\author[0000-0002-4410-5387]{A.~Rest}
\affiliation{Department of Physics and Astronomy, The Johns Hopkins University, Baltimore, MD 21218, USA}
\affiliation{Space Telescope Science Institute, Baltimore, MD 21218, USA}
\email{arest@stsci.edu}

\author[0000-0003-2445-3891]{M.~R.~Siebert}
\affiliation{Space Telescope Science Institute, Baltimore, MD 21218, USA}
\email{msiebert@stsci.edu}

\author[0000-0002-0840-6940]{S.~K.~Yadavalli}
\affiliation{Center for Astrophysics | Harvard \& Smithsonian, 60 Garden Street, Cambridge, MA 02138-1516, USA}
\email{karthik.yadavalli@cfa.harvard.edu}

\author[0000-0002-0632-8897]{Y.~Zenati}
\affiliation{Department of Physics and Astronomy, The Johns Hopkins University, Baltimore, MD 21218, USA}
\affiliation{Space Telescope Science Institute, Baltimore, MD 21218, USA}
\affiliation{Astrophysics Research Center of the Open University (ARCO), The Open University of Israel, Ra’anana 4353701, Israel}
\affiliation{Department of Natural Sciences, The Open University of Israel, Ra’anana 4353701, Israel}
\email{yzenati1@jhu.edu}

\author[0000-0001-5233-6989]{Q.~Wang}
\affiliation{Department of Physics and Kavli Institute for Astrophysics and Space Research, Massachusetts Institute of Technology, 77 Massachusetts Avenue, Cambridge, MA 02139, USA}
\affiliation{TESS-ULTRASAT Joint Postdoctoral Fellow}
\email{qnwang@mit.edu}

\begin{abstract}
In recent years, multiple Type Ia supernovae (SNe\,Ia) have been observed with ``bumps'' in their rising light curves shortly after explosion. Here, we present SN\,2021qvo: a SN\,Ia that exhibits a clear early bump in photometry obtained by the Young Supernova Experiment. 
Photometric and spectroscopic observations of SN\,2021qvo show that it has a broader light curve, higher peak luminosity, shallower \SiII\ $\lambda$5972 pseudo-equivalent width, and lower ejecta velocities than normal SNe\,Ia, which are all consistent with the characteristics of the 2003fg-like (often called ``super-Chandrasekhar") SN subtype. Including SN~2021qvo, just four known 2003fg-like SNe\,Ia have sufficient pre-peak data to reveal a rising light-curve bump, and all four have bump detections.  
Host-galaxy analysis reveals that SN\,2021qvo exploded in a low-mass galaxy ${\rm log}(M_{\ast}/M_{\odot}) = 7.83^{+0.17}_{-0.24}$, also consistent with other members of this class.
The current leading early-bump 2003fg-like SN\,Ia progenitor model involves an interaction between the circumstellar material (CSM) and the SN ejecta. We test the validity of this theory by modeling the early bump and subsequent light-curve evolution of SN\,2021qvo with the Modular Open Source Fitter for Transients.
We find that the bump can be modeled with a best-fit CSM mass in the range $M_\mathrm{CSM}=3.31-8.51 \times 10^{-3} M_\odot$.
SN\,2021qvo adds to the small but growing number of 2003fg-like SNe\,Ia with rising light-curve bumps; as the number of these SNe\,Ia with CSM estimates continues to grow, population-level inferences about the CSM distribution will be able to constrain the progenitor scenario for these SNe\,Ia. 
\end{abstract}

\keywords{Type Ia supernovae(1728)}

\section{Introduction} \label{sec:intro}

Type Ia supernovae (SNe\,Ia) follow a strongly correlated luminosity-width relation (LWR) that relates the light curve shape of a SN\,Ia to its intrinsic luminosity, allowing for distances to be calculated with high precision \citep{Phillips93}. 
SNe\,Ia can be used to measure cosmological parameters with high precision using distances derived from this relationship, including the Hubble constant \citep[e.g.,][]{Burns_2018, Freedman19, Riess22, Freedman_2024} and the dark energy equation of state parameter \citep[e.g.,][]{Betoule14, Scolnic18, Jones19, Brout22, descollaboration2024dark}.

While there is broad consensus that SNe\,Ia result from an exploding Carbon-Oxygen White Dwarf (CO WD) star \citep{Hoyle60}, there is a large diversity of proposed models explaining SNe\,Ia, each with their own unique progenitor scenario and explosion mechanism (see \citealp{Ruiter25} for a recent review). SNe\,Ia progenitor channels include the single degenerate scenario containing a CO WD and a non-degenerate companion \citep[e.g.,][]{Whelan73, Nomoto82a}, the double degenerate scenario containing two degenerate CO WDs, a CO WD and a He WD, or a CO WD and a HeCO WD \citep[e.g.,][]{Nomoto80, Iben84, Perets+19_Ia}, and the core degenerate scenario containing a CO WD and an asymptotic giant branch star \citep[e.g.,][]{Kashi11}.

Each progenitor scenario has several possible explosion mechanisms that could result in a SN\,Ia. 
In the single degenerate scenario, the CO WD accretes material from the non-degenerate companion star (either a red giant star or a main sequence star) until a thermonuclear runaway explosion occurs \citep[generally when the CO WD nears the Chandrasekhar mass;][]{Whelan73, Nomoto84}. Accretion-triggered explosions can also occur in the double degenerate scenario \citep[e.g.,][]{Nomoto82a}; a detonation of accreting or built-up surface helium may send a converging shock into the CO WD, causing central carbon ignition \citep[and thus a SN\,Ia, e.g.,][]{Nomoto80, Nomoto82b, Livne90, Shen14b}. Finally, dynamical or violent mergers may also cause a SN\,Ia in the double degenerate scenario \citep[e.g.,][]{Iben84, Webbink84, Rosswog09, Raskin09, Pakmor10, Kwok_2023a, Kwok_2024, Siebert_2024}.

Most SNe\,Ia are spectroscopically and photometrically ``normal'' \citep[e.g.,][]{Desai_2024} with similar photometric and spectroscopic features \citep[e.g.,][although at the earliest stages show significant diversity, \citealp{Hoogendam_2025a}]{Morrell_2024}. However, a growing number of SNe\,Ia have been discovered with different observational properties, which may arise from differences in the underlying progenitor scenario and/or explosion mechanism. While some SN\,Ia subtypes follow a luminosity-width relationship (i.e., 1991bg-like and 1991T-like SNe\,Ia,  \citealp{Filippenko_91T, Leibundgut93,Phillips92, Filippenko_91T}), other subtypes do not.  These include 2002cx-like \citep{Li_2003}, 2002es-like \citep{Ganeshalingam_2012}, and 2003fg~like SNe\,Ia \citep{Howell_03fg}. Because of deviations in parameters like peak absolute magnitude and light-curve decline rate, such SNe\,Ia are not always standardizable for SN cosmology. Still, these peculiar objects offer an opportunity to understand the extreme ends of the SNe\,Ia population.

These peculiar SN\, Ia subtypes are identified through a mix of photometric and spectroscopic characteristics. Several 2003fg-like SNe Ia have abnormally luminous abs mags between $-$19.5 to $-$20.4 \citep{Hicken_2007, Scalzo10, Dimitriadis22} and decline rates $\Delta m_{15}$ values in the range of 0.7 to 0.9~mag \citep{Taubenberger_2017, Ashall_2021}, however we now know there is a broad diversity in this subclass with some showing more typical luminosities and decline rates ($\sim$$0$-$1$ mag brighter \citealp{Hsiao_2020, Lu_2021}).
In general, 2003fg-like events are $\sim$$1$-$2$ mag brighter than normal SNe\,Ia \citep{Desai_2024} and have light curve shapes consistent with or broader than the slowest-declining normal SNe\,Ia.
With these properties in mind, common photometric diagnostic features are a primary \textit{i}-band maximum after the primary \textit{B}-band maximum, weak or missing \textit{i}-band secondary maximum, and a broad light curve \citep[although the latter two are seen in only some 2003fg-like SNe\,Ia]{Ashall_2021}.

Spectroscopically, 2003fg-like SNe\,Ia also exhibit different features than normal SNe\,Ia. The early-time spectra of 2003fg-like SNe\,Ia have weaker \CaII\ features and prominent \OI\ and \CII\ features \citep{Taubenberger_2017, Ashall_2021} and have higher UV flux than normal SNe\,Ia \citep{Hoogendam_2024,Bhattacharjee_2025}. In general, the \CII~$\lambda\lambda$6580,7234 features are especially strong for up to $\sim$2 weeks after maximum light. At later times, 2-4 weeks after maximum light, 2003fg-like SN\,Ia spectra are dominated by \FeII\, slightly earlier than for normal SNe\,Ia \citep{Silverman11, Taubenberger11, Chakradhari14, Parrent16}. Other spectroscopic diagnostics to identify a SNe as 2003fg-like include a lack of an \textit{H}-band break at +10 days in the near-infrared spectra and a low ionization state in nebular-phase spectra. Features common, but not ubiquitous in the 2003fg-like population include strong \CII\ features after maximum light, low ejecta velocity gradients before maximum light, a lack of \TiII\ lines in the peak time spectra, and asymmetric near-infrared nebular lines \citep[e.g, SN\,2022pul;][]{Kwok_2024, Siebert_2024, OHora_2025}.

Recently, three 2003fg-like SNe\,Ia with sufficiently early detections have been discovered with bump-like features in their early-time emission:\footnote{Following \citet{Hoogendam_2024}, we use the term ``bump" to indicate a non-monotonic rise in the SN light curve.} SNe\,2020hvf \citep{Jiang21}, 2021zny \citep{Dimitriadis23}, and 2022ilv \citep{Srivastav23b}. These three SNe\,Ia are the only three known 2003fg-like SNe\,Ia with sufficient observations to ascertain the presence of a rising light curve bump, suggesting these bump features may be common or even ubiquitous in 2003fg-like SNe\,Ia. Furthermore, these bumps have solely been observed in 2003fg-like SNe\,Ia and in another SN\,Ia subtype, 2002es-like SNe\,Ia, potentially suggesting these objects are linked \citep{Hoogendam_2024}.

Many models attempt to explain the origin of rising light curve bumps in SNe\,Ia. While the interaction between SN ejecta and a non-degenerate companion in the single degenerate scenario \citep[e.g,][]{Kasen_2010, Maeda_2014, Kutsuna15} has been ruled out for these objects \citep{Dimitriadis23,Srivastav23b,Hoogendam_2024}, varying radial distributions of $^{56}\rm{Ni}$ \citep[e.g.,][]{Piro_2016, Magee20b}, the interaction between SN ejecta and circumstellar material (CSM; e.g., \citealp{Levanon15, Piro_2016, Levanon17, Maeda23}), or a surface He detonation \citep[e.g.,][]{Jiang17, Maeda_2018, Polin_2019, Leung_2020, Leung_2021} may produce an observable ``bump'' signature. The early colors may be too blue for significant surface $^{56}\rm{Ni}$ from either mixing or a surface He-detonation, leaving CSM interaction as the preferred explanation \citep{Hoogendam_2024}. This is consistent with the spectral and light curve evolution, which also suggests a CSM-interacting explosion \citep{Ashall_2021}. 

In this work, we examine the photometric and spectroscopic properties of SN\,2021qvo, a 2003fg-like SN\,Ia that adds to the class of such objects that exhibit early light-curve bumps. Section \ref{sec:data} presents our observations. Sections \ref{sec:photometry} and \ref{sec:spectroscopy} contain our photometric and spectroscopic observations, respectively. In Section \ref{sec:discussion}, we constrain the CSM mass and compare to similar events in the literature. Finally, Section \ref{sec:summary} summarizes the main results. Throughout, we use a flat $\Lambda$CDM cosmology with $\Omega_m = 0.3$ and H$_0 = 70~{\rm km\ s^{-1}\ Mpc^{-1}}$.

\section{Data} \label{sec:data}

SN\,2021qvo was discovered in a small, faint galaxy (see Section \ref{sec:host-galaxy}) by the Asteroid Terrestrial-impact Last Alert System (ATLAS; \citealp{Tonry18}) on June 21, 2021  with an $o$-band magnitude of $\sim$19.5~mag \citep{ATLAS_discovery}. The location of SN\,2021qvo is $(\alpha,\delta)~=~(22^{\mathrm{h}}08^{\mathrm{m}}12\fs08,+07\arcdeg07\arcmin00\farcs78)$

The Young Supernova Experiment (YSE; \citealp{Jones21}) classified SN\,2021qvo two days after its discovery on June 23, 2021 \citep{21qvo_Classification} with a classification spectrum from the Alhambra Faint Object Spectrograph and Camera on the Nordic Optical Telescope. 
They used \texttt{SuperFit} \citep{Howell05} and SNID \citep{SNID} to classify the event as a SN\,Ia, finding that the transient matched the template of a SN\,Ia around $-10$~days.

\subsection{Photometric Data}
Photometric observations of SN~2021qvo primarily include the ATLAS ``cyan" and ``orange" filters (approximately equivalent to $g+r$ and $r+i$ bandpasses, respectively) and the Pan-STARRS $griz$ filters, spanning from MJD $\sim$59346 to $\sim$59880.
We obtained the ATLAS photometry from the online ATLAS forced photometry service \citep{smith20,Shingles21}.\footnote{\href{https://fallingstar-data.com/forcedphot/}{https://fallingstar-data.com/forcedphot}.} The ATLAS forced photometry has a detection of SN\,2021qvo before discovery on 17 June at 4.7$\sigma$ significance.

 \begin{deluxetable}{cccc}
    \tablecaption{Photometric Measurements of SN\,2021qvo from ATLAS.
    \label{tab:photometry_data_ATLAS}}
    \tablehead{ \colhead{MJD} & \colhead{Filter} & \colhead{Magnitude} & \colhead{Mag err}
    }
    \startdata
    58402.41 &  c &   21.24 &   0.60  \\
    58406.37 &  o &   21.01 &   0.53  \\
    58410.36 &  o &   18.09 &   1.09  \\
    \vdots & \vdots & \vdots & \vdots \\ 
    59592.21 &  o &   19.86 &   0.59  \\
    \enddata
    \tablecomments{The full table will be available in the online journal.}
\end{deluxetable}

\begin{deluxetable}{cccc}
    \tablecaption{Photometric Measurements of SN\,2021qvo from YSE.
    \label{tab:photometry_data_YSE}}
    \tablehead{ \colhead{MJD} & \colhead{Filter} & \colhead{Magnitude} &     \colhead{$\sigma_m$}
              }
    \startdata
        59382.57 & $g$ &  20.04 & 0.07 \\
        59382.57 & $i$ &  20.25 & 0.08 \\
        59383.57 & $g$ &  20.50 & 0.08 \\
        \vdots & \vdots & \vdots & \vdots \\
        59527.21 & $i$ &  21.24 & 0.23 \\
    \enddata
    \tablecomments{The full table will be available in the online journal.}
\end{deluxetable}

Before the ATLAS discovery, serendipitous pre-discovery observations were taken by YSE with the Pan-STARRS Giga Pixel Camera 1 \citep[GPC1;][]{PanStarrs_2002, PanStarrs_2007}. YSE is a time-domain survey using the Pan-STARRS telescopes designed to obtain $griz$ light curves of transients up to $z=0.2$. YSE can detect events as faint as $\sim$20.5~mag in $z$ and $\sim$21.5~mag in $gri$, observing 1500 deg$^2$ every 3 days \citep{Jones21}. The Pan-STARRS photometry used in this manuscript has already been published in YSE DR1 \citep{Aleo23}. As described in that data release, the Pan-STARRS photometric data collected by YSE is processed by the University of Hawai`i Institute for Astronomy's Image Processing Pipeline \citep{Magnier_2020a, Magnier_2020b}.  The first detection by YSE was on 17 June 2021 in the $g$ and $i$ bands, approximately 17 days before maximum light and $\sim$4~days before the ATLAS discovery. After discovery, we used the YSE-PZ interface \citep{Coulter23} to triage the data and organize follow-up observations of this event. See Tables \ref{tab:photometry_data_ATLAS} and \ref{tab:photometry_data_YSE} for the ATLAS and Pan-STARRS photometric data respectively.

Beginning near maximum light, imaging of SN~2021qvo was also obtained in the $BVri$ bands with the 1\,m Nickel telescope at Lick Observatory. The images were calibrated using bias and sky flat-field frames following standard procedures. PSF photometry was performed, and photometry was calibrated using Pan-STARRS photometric standards \citep{Flewelling20}.

\subsection{Hubble Space Telescope Data}
\label{sec:hst}

The location of SN\,2021qvo was also imaged by the {\it Hubble Space Telescope} ({\it HST}; GO 16691) on MJD~59841, approximately 440 days after $g$-band maximum, in the F555W and F850LP filters. To see if these data could place meaningful constraints on the presence of CSM interacting with the SN ejecta at late times (i.e., \citealp{Graham19}), we aligned the {\it HST} imaging to Pan-STARRS $r$ using 13 stars visible in all images. Our astrometric alignment procedures followed those of \citet{Angus24}, including star detection via Source Extractor \citep{Bertin96} to identify common stars in all images. Specifically, we removed sources within 20 pixels of the edges of the images, required sources to have ellipticity $< 0.5$ and FWHM less than five times the mode of the FWHM distribution of detected sources, in order to exclude extended or irregular objects. After applying these criteria, we visually confirmed 13 high-confidence point sources common to both the {\it HST} and Pan-STARRS images and used these for the transformation. We performed iterative fitting to determine the best astrometric solution with the {\tt IRAF } tool \texttt{GEOMAP}, and used {\tt IRAF}'s \texttt{GEOTRAN} to apply the resulting transformation \citep{Tody86}. The final uncertainty on the position was approximately 2 HST pixels. We selected a region of the environment and calculated the probability that a chance alignment occurs within a 2-pixel radius aperture to be 0.01\%, indicating that the source is unlikely to be a chance coincidence.

Unfortunately, there was no unambiguous detection of SN\,2021qvo, although there was a possible detection in F555W within the uncertainty interval of the best-fit position. Because SN\,2021qvo is within the isophotal radius of its host galaxy, we measured the photometry from the cosmic ray-corrected images within a 5-pixel radius using zero points and aperture corrections provided by the WFC3 team.\footnote{\url{https://www.stsci.edu/hst/instrumentation/wfc3/data-analysis/photometric-calibration/uvis-photometric-calibration}.} For sky subtraction, we used the {\tt photutils} package from \texttt{AstroPy} to find pixels along the same elliptical isophote relative to the host-galaxy center of SN\,2021qvo; we constructed an annulus around the host galaxy with a 10-pixel width to match the diameter of the photometric aperture.

Due to the lack of a galaxy template, we cannot be confident that we have detected the SN rather than galaxy flux; the measured magnitude of ${\it F555W} = 26.25 \pm 0.28$~mag (AB) is insufficient to provide meaningful late-time CSM interaction constraints. 

\begin{deluxetable}{lrrrr}
\tablecaption{Spectroscopic observations of SN~2021qvo.\label{tab:spec}}
\tablewidth{0pt}
\tablehead{
\colhead{UT Date} & \colhead{MJD} & \colhead{Epoch\tablenotemark{\rm a}} &  \colhead{Telescope} & \colhead{Instrument} \\ & \colhead{[days]} & \colhead{[days]} & & }
\startdata
2021-06-23 &    59388.00    & $-13.2$   & NOT       & ALFOSC    \\
2021-07-09 &    59404.40    & $  3.2$   & Lick      & KAST      \\
2021-07-15 &    59410.16    & $  9.0$   & Lick      & KAST      \\
2021-07-19 &    59414.82    & $ 13.6$   & Goodman   & HST      \\
2021-10-14 &    59501.01    & $ 99.8$   & Gemini-S  & GMOS      \\
2021-10-14 &    59501.06    & $ 99.9$   & Gemini-S  & GMOS      \\
2021-11-06 &    59524.00    & $122.8$   & Keck I    & LRIS      \\
2025-07-27$^{\rm b}$ &    60883.40    & $1482.2$  & Keck I    & LRIS      \\
\enddata

\tablenotemark{\rm a}{Phase relative to rest frame $B$-band maximum on MJD 59401.2.}

\tablenotemark{\rm b}{Host-galaxy light only.} 

\end{deluxetable}

\subsection{Spectroscopic Data and Redshift Determination}

SN\,2021qvo has seven epochs of spectroscopic data spanning $\sim$$-13$~days to $\sim$$+123$~days relative to maximum light (Table \ref{tab:spec}).  Spectra were taken with ALFOSC, the Kast Double Spectrograph \citep{Miller94}, the Goodman High Throughput Spectrograph \citep{Goodman}, the Gemini Multi-Object Spectrograph \citep{GMOS}, and the Low-Resolution Imaging Spectrometer \citep{LRIS}. An additional host-galaxy spectrum was taken with LRIS on 2025-07-27. Table \ref{tab:spec} logs the spectroscopic data for SN\,2021qvo. The Kast and LRIS data were reduced using the UCSC Spectral Pipeline.\footnote{\url{https://ucsc-spectral-pipeline.readthedocs.io/en/latest/}.} All spectra were corrected for bias, overscan, and flat-field frames. Wavelength calibrations used arc lamps acquired during the same night and the same instrument settings. When possible, the telluric bands were removed using standard star observations.

To determine the redshift from the final host-galaxy spectrum, we use the the {\tt XCSAO} package \citep{Kurtz92}.  {\tt XCSAO} compares spectra to a set of galaxy spectral templates and measures the best-fit cross-correlation redshift; from the host-galaxy spectrum of SN~2021qvo, we find a precise redshift of 0.04205.

\section{Host-galaxy Properties}\label{sec:host-galaxy}

\begin{figure*}[t]
    \centering
    \includegraphics[width=\linewidth]{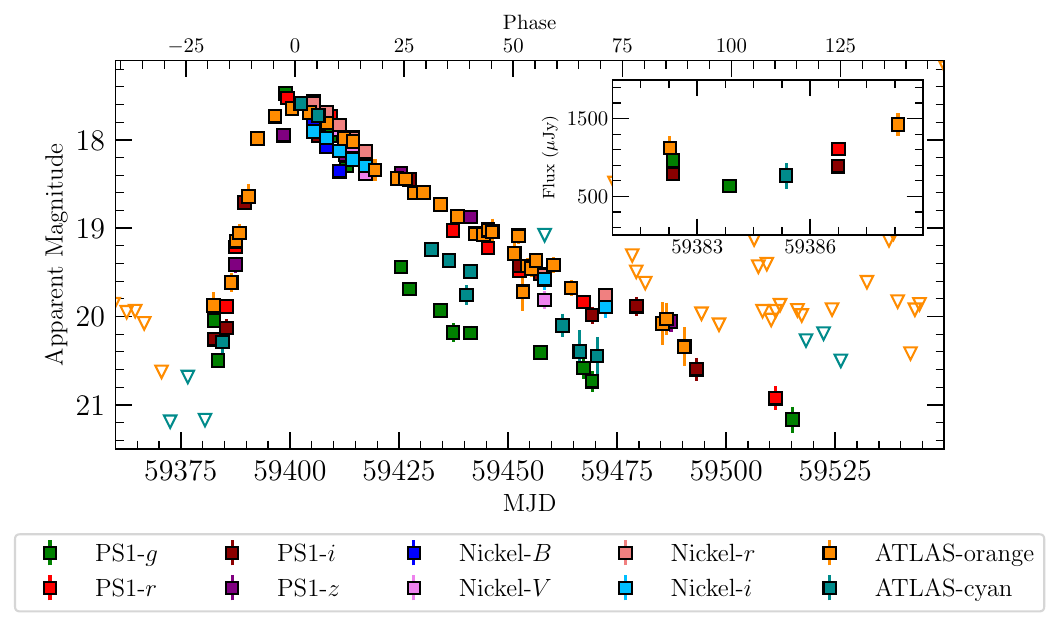}
    \caption{SN\,2021qvo observations in different filters, plotted as days from the $g$-band peak (MJD=59401.2) versus the observed magnitude. Downward-facing triangles denote upper limits. The inset shows a zoomed-in portion of the light curve in flux space from $-20$ to $-15$ days before peak, showing the early time light-curve bump.}
    \label{fig:zoomlc}
\end{figure*}

\begin{figure}
    \centering
    \includegraphics[width=\linewidth]{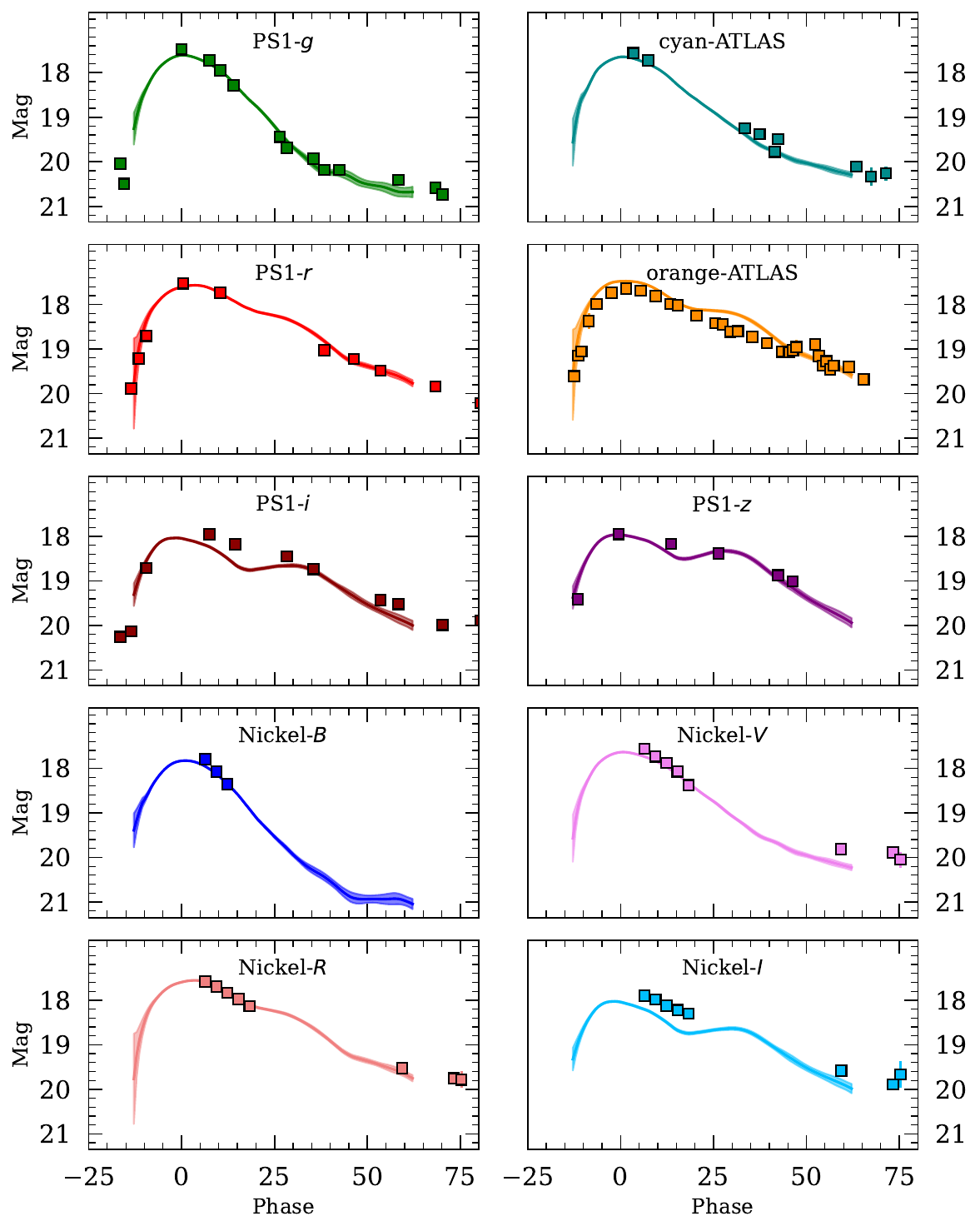}
    \caption{\texttt{SNooPy} fits for all ten bands of SN\,2021qvo. Plotted as time from $B$-band peak given by \texttt{SNooPy} versus the observed magnitude. Upper limits were removed when fitting the light curve.}
    \label{fig:snoopy}
\end{figure}

We estimated the host-galaxy properties of SN\,2021qvo using the {\tt Blast} web application \citep{Jones24},\footnote{\url{https://blast.scimma.org/} with 2021qvo's results at \url{https://blast.scimma.org/transients/2021qvo/}.}. {\tt Blast} estimates physical parameters of transient host galaxies by measuring photometry from GALEX \citep{Martin05_GALEX}, SDSS \citep{York00_SDSS, Blanton_2017SDSS, Ahumada20_SDSS}, Pan-STARRS \citep{Chambers16_PanSTARRS}, DECam \citep{Dey19_DESI}, 2MASS \citep{Skrutskie_2MASS}, and WISE \citep{Wright10_WISE}.  {\tt Blast} uses global host-galaxy apertures that are adjusted for the point-spread function (PSF) of each individual filter/instrument to measure the photometry and fits to stellar population synthesis models via the Prospector-$\alpha$ model \citep{Leja17, Johnson21}, which is implemented via the SBI$++$ neural posterior estimator for increased speed \citep{Wang23}. 

The host galaxy of SN\,2021qvo is faint, with PS1 $g \simeq 22.1$~mag, and is nearly coincident with the SN location, at ($\alpha$, $\delta$) of (22:08:12.11, $+$07:07:00.84). SN\,2021qvo occurred in an unusually faint, low-mass host galaxy with a stellar mass ${\rm log}(M_{\ast}/M_{\odot}) = 7.83^{+0.17}_{-0.24}$~dex. Although this is less massive even than most reported 2003fg-like hosts, which prefer low-mass galaxies \citep{Ashall_2021,Lu_2021}, the host galaxy of at least one --- SN~2022ilv --- is even fainter \citep{Srivastav23a}.  {\tt Blast} also reports host-galaxy masses ${\rm log}(M_{\ast}/M_{\odot}) < 9$~dex for 2003fg-like SNe 2007if and iPTF13asv \citep[see also][]{Scalzo10,Cao2016}.  While not highly star-forming, there is an indication of recent star formation in the host of SN~2021qvo, with a global star formation rate of ${\rm log}_{10}({\rm SFR\ M_{\odot}/yr}) = -1.82^{+0.59}_{-0.78}$ 
and a global specific star formation rate ${\rm log}({\rm sSFR\ yr^{-1}}) = -9.70	^{+0.82}_{-0.79}$. The host-galaxy dust is not well constrained by {\tt Blast}, with best-fit $A_V$ consistent with $\sim$1-4 magnitudes of extinction.  The stellar metallicity is poorly constrained by the broadband photometry alone, but {\tt Prospector} finds modest evidence that the host has a low metallicity, ${\rm log}(Z_{\ast}/Z_{\odot}) = -1.20^{+0.64}_{-0.47}$, which is typical of 2003fg-like SNe\,Ia and may imply low-metallicity progenitors \citep[e.g.,][]{Childress11, Hsiao_2020, Lu_2021, Ashall_2021}.

 \begin{figure}
    \centering
    \includegraphics[width=\linewidth]{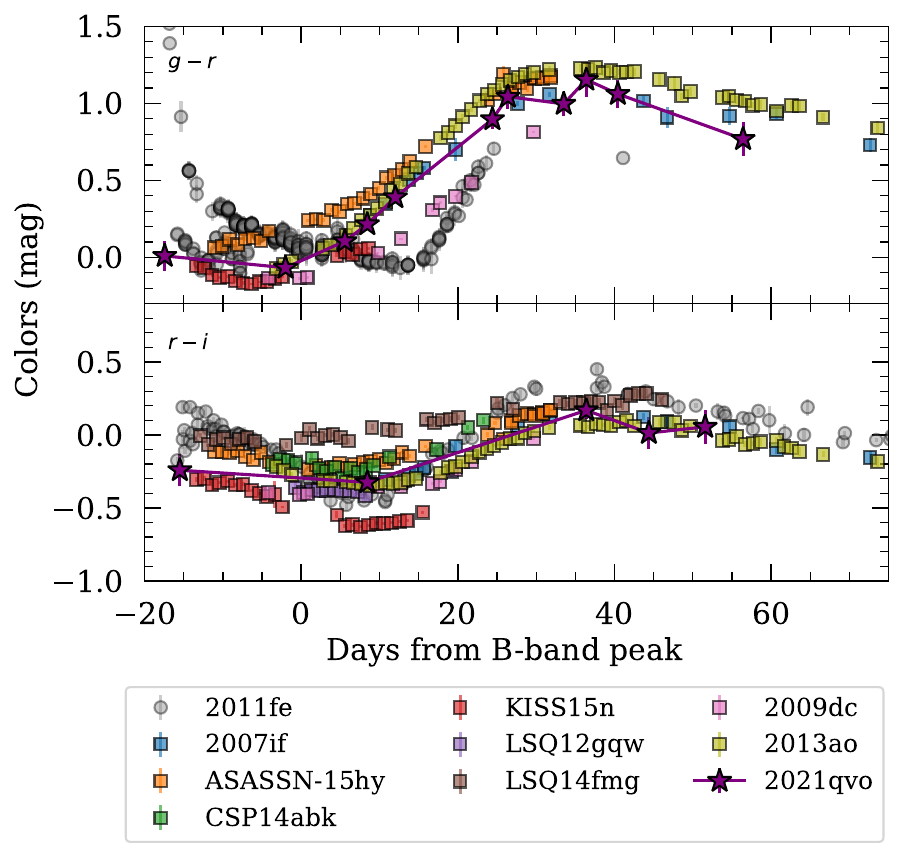}
    \caption{Color evolution plot for SN\,2021qvo in the $\it{r-i}$ and $\it{g-r}$ bands. The data used are from the $g$, $r$, and $i$ bands collected by Pan-STARRS. A comparison is made using other confirmed 2003fg-like SNe\,Ia derived from the Carnegie Supernova Project \citep{Ashall_2021}, as well as a normal type SN\,Ia SN\,2011fe. Curves were not corrected for host-galaxy extinction.}
    \label{fig:color-curve}
\end{figure}

Compared to the larger population of SNe, SN\,2021qvo is relatively unique.  It has a mass lower than 98.7\% of SN\,Ia hosts measured by {\tt Blast}, and 99.5\% of transients overall; the comparison set includes 5408 spectroscopically classified SNe, of which 3943 are SNe\,Ia and eight have been classified as 2003fg-like.
Its sSFR is somewhat less extreme; it is higher than 84\% of SNe\,Ia and 81\% of all transients.
 Lastly, SN\,2021qvo's host-galaxy metallicity is lower than 87\% of transient hosts measured by {\tt Blast}, and 89\% of SNe\,Ia.\footnote{The comparison sample of {\tt Blast} hosts is primarily sourced from magnitude-limited Transient Name Server discoveries, and we do not rigorously evaluate the selection effects in this work.  If we restrict to blindly-selected TNS-sourced transients, the results are nearly identical. While we we have not fully vetted the individual {\tt Blast} results for this comparison, the overall trends are reliable.}

\section{Photometry} \label{sec:photometry}
In this section, we discuss the photometric properties of SN\,2021qvo, including the light curve, the color-stretch parameter $s_{BV}$, the color curves, and the peak magnitudes, and compare them to other SN\,Ia subtypes. This analysis will help characterize the properties of SN\,2021qvo and solidify it as a 2003fg-like SN event. The photometric data for SN\,2021qvo are summarized in Tables \ref{tab:photometry_data_ATLAS} and \ref{tab:photometry_data_YSE}.

\subsection{Light Curves}

Figure \ref{fig:zoomlc} shows the light curve of SN\,2021qvo, with the inset showing the rising light curve bump around 18.5~days before peak light. The bump is most clearly visible in the $g$ band, with a $\sim4.4\sigma$ detection of the decrease in flux. The $i$ band also shows a signature attributable to a bump: a nearly flat light-curve evolution over a span of $\sim$3~days nearly 2~weeks before maximum light is highly atypical of all known SNe\,Ia. A natural explanation for the flat $i$-band light curve is that SN~2021qvo rises through or peaks at the first $i$-band detection, declines, and again rises through the second $i$-band detection. Alternatively, the light curve could be declining through the second point, but the later evolution requires it to rise again. Finally, the $o$-band light curve also exhibits a 4.7$\sigma$ flux excess. Given the multi-band signatures, especially in the $g$ band, which is well above the YSE detection threshold, we conclude there is a highly significant rising light curve bump in SN~2021qvo. 

We next use the SuperNova in object-oriented Python (\texttt{SNooPy}) software \citep{Burns_2011, Burns_2014} to determine the light-curve parameters of SN\,2021qvo. The \texttt{SNooPy} package is designed to fit SN\,Ia photometry to an empirical light-curve model. Specifically, we use \texttt{SNooPy}'s \texttt{EBV\_model2}, which fits the supernova magnitudes, time of $B$-band maximum light, host galaxy extinction, and a light-curve shape parameter that can be either the ``color--stretch" parameter $s_{BV}$, the time difference between the $B$-band maximum and the reddest point in the ($B-V$) color curve normalized by 30 days \citep{Burns_2014}, or $\Delta m_{15}(B)$, the difference in $B$-band magnitude between peak and 15 days after peak \citep{Phillips93}.  Here, we use $s_{BV}$, which is a more recent and improved stretch parameter \citep{Burns_2014}; it also allows easy comparison to measurements from previous 2003fg-like samples \citep{Ashall_2021}. The \texttt{SNooPy} fit to SN\,2021qvo is shown in Figure \ref{fig:snoopy}; note that the \texttt{SNooPy} model assumes SN\,2021qvo is a normal SN\,Ia and therefore predicts a secondary maximum in the $i$ and $z$ bands; however, secondary $i$- and $z$-band maxima are not present in 2003fg-like SNe\,Ia \citep{Taubenberger_2017, Ashall_2021}.

To minimize errors resulting from the lack of 2003fg-like SN\,Ia fitting templates in \texttt{SNooPy}, we first used the built-in Gaussian process function to fit all our available bands with the early time bump masked out. We use these fits to set priors on the time of maximum light in the subsequent {\tt EBV\_model2} fits. SN\,2021qvo peaks in the $B$-band\footnote{While we do not have $B$-band photometry, {\tt SNooPy} determines the time of peak in the $B$-band by mangling the \citet{Hsiao07} SED to match the colors of SN\,2021qvo.  It then computes synthetic $B$-band light curves from the mangled SED.} on $MJD = 59401.2 \pm 1.00 $~days at $m_{g,max} = 17.42 \pm 0.03$~mag and has a color stretch parameter $s_{gr} = 1.210 \pm 0.116$ (using the linear correlation found by \cite{Ashall_2020} we calculate an $s_{BV} = 1.184 \pm 0.121$). Correcting only for the redshift-derived distance, we estimate a $g$-band peak absolute magnitude of $M_{g,max} = -19.3\pm0.249$ mag. This ranks SN\,2021qvo among the lower-luminosity 2003fg-like SNe\,Ia such as SN\,2012dn \citep{Chakradhari14} and SN\,2022pul \citep{Siebert_2024}.

\subsection{Color Curves}

Figure \ref{fig:color-curve} shows the $g-r$ and $r-i$ color curves for SN\,2021qvo and a comparison sample of other 2003fg-like SNe\,Ia from the Carnegie Supernova Project corrected for Milky-Way extinction but without host-galaxy extinction correction. The comparison sample includes SNe\,2007if \citep{Akerlof_2007}, 2009dc \citep{Puckett_2009}, LSQ12gqw \citep{Baltay_2013a}, 2013ao \citep{Ashall_2021}, LSQ14fmg \citep{Baltay_2013a}, CSP14abk \citep{Ashall_2021}, ASASSN-15hy \citep{Holoien_2015}, and KISS15n \citep{Ashall_2021}. 

SN\,2021qvo shows a color evolution consistent with other 2003fg-like SNe\,Ia. As detailed by \citet{Ashall_2021}, the $r-i$ curve demonstrates the greatest differences between individual 2003fg-like events in terms of their bluest point and color curve shape. Although the data are somewhat limited, SN\,2021qvo is bluer in $r-i$ near peak and appears to have larger color changes as a function of phase than other 2003fg-like SNe\,Ia, with the largest color difference being $\Delta (r-i) = 0.49$~mag between the bluest and reddest point and a $\Delta (r-i) = 0.11$~mag within 15 days at later epochs. This leads to a bluest point around 8 days post $B$-band peak at $r-i = -0.33$~mag, similar to the bluest range of most other 2003fg-like SNe\,Ia and redder than normals (see Figure 9 in \citealp{Ashall_2021}). We then see a flattening after the reddest point around $\sim$40 days until $\sim$80 days, consistent with other 2003fg-like events. After this, the large uncertainties make inferring the color evolution of SN\,2021qvo tenuous.

\begin{figure}
    \centering
    \includegraphics[width=\linewidth]{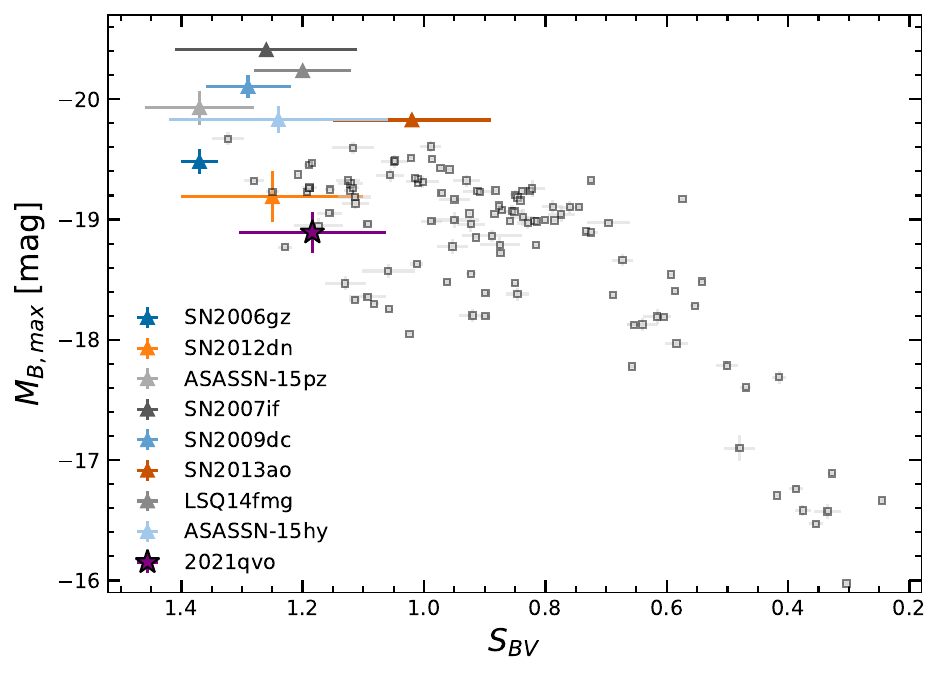}
    \caption{Luminosity Width Relation, $i.e.$ Phillips relation \citep{Phillips_1999}, plot for SN\,2021qvo and multiple 2003fg-like events, showing the relationship between the color stretch parameter $s_{BV}$ and the absolute magnitude in the $B$-band $M_{B,max}$. The background sample of normal SNe\,Ia (grey squares) is taken from \citet{Burns_2018}. SN\,2021qvo is shown as a purple star. We increased the error in SN\,2021qvo's $M_{B,max}$ to account for the sources of underluminosity detailed in Section \ref{subsection:LWR}. All other colored triangles are confirmed 2003fg-like events that have measurements in the $B$-band. These events have been corrected for Milky-Way extinction.}
    \label{fig:lwr}
\end{figure}

The $g-r$ curve shows a pre-peak evolution consistent with the early colors of KISS15n, starting around $g-r=0.01$~mag before reaching its bluest point $\sim$2 days before peak. The curve then becomes redder like the other 2003fg-like SNe\,Ia in an almost monotonic nature. After reaching its reddest point at around 40 days, SN\,2021qvo gradually becomes bluer at later times. This continues until 80 days, when again, the large uncertainties prevent robust conclusions about the color evolution at later phases.

\subsection{Luminosity Width Relation}
\label{subsection:LWR}

The LWR \citep{Phillips93} relates the light curve shape of a SN\,Ia to its intrinsic luminosity. Normal SNe\,Ia have a tightly correlated LWR, with the more luminous SNe\,Ia having slower decline rates, and the less luminous SNe\,Ia having faster decline rates. 2003fg-like SNe\,Ia are often located on the luminous and broad area of the LWR parameter space. However, 2003fg-like SNe\,Ia span a variety of luminosities and light curve shapes, including some extremely luminous events (e.g., SN~2007if, \citealp{Scalzo10}; SN~2009dc, \citealp{Taubenberger11, Silverman11}; and LSQ14fmg, \citealp{Hsiao_2020}) and some underluminous events (relative to their light curve shape, e.g., ASASSN-15hy \citealp{Lu_2021}).

\begin{figure}
    \centering
    \includegraphics[width=\columnwidth]{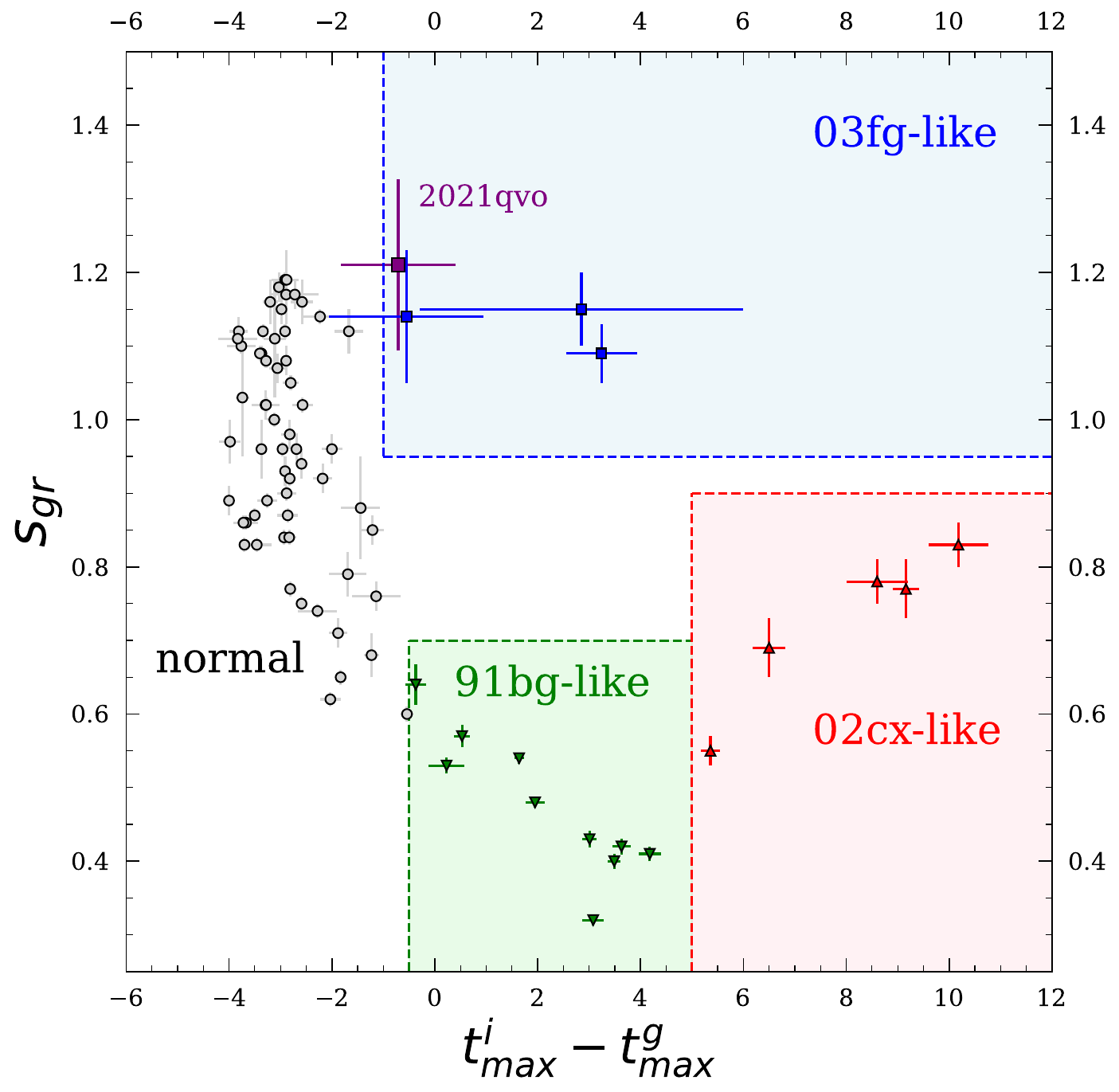}
    \caption{Color--stretch parameter $s_{gr}$ as a function of the $g$-band maximum relative to the $i$-band maximum. SN\,2021qvo (purple) can be seen in the top right. Approximate boxes are laid out to coincide with the classifications of different SN\,Ia subtypes \citep[2003fg-like, 1991bg-like, and 2002cx-like SN\,Ia]{Ashall_2020}. Diagram derived from Figure 3 in \textcite{Ashall_2020}.}
    \label{fig:ashall_diagram}
\end{figure}

Figure \ref{fig:lwr} shows SN\,2021qvo on the $B$-band vs $s_{BV}$ LWR diagram. We compare SN\,2021qvo to normal SNe\,Ia from CSP-I \citep{Krisciunas17, Burns_2018} and confirmed 2003fg-like SNe\,Ia from \citep{Ashall_2021}. The 2003fg-like sample includes SNe\,2006gz \citep{Hicken_2007}, 2007if \citep{Akerlof_2007}, 2009dc \citep{Puckett_2009}, 2012dn \citep{Chakradhari14}, 2013ao \citep{Ashall_2021}, LSQ14fmg \citep{Baltay_2013a}, ASASSN-15hy \citep{Holoien_2015}, and ASASSN-15pz \citep{Chen_2019}. Although it is the least luminous object in the sample, the location of SN\,2021qvo on the LWR is coincident with the known sample of 2003fg-like SNe. SN\,2021qvo is a notably underluminous 2003fg-like, instead blending into the normal SN\,Ia population rather than some of the more overluminous 2003fg-like SNe\,Ia. We note, however, a couple of reasons for this underluminosity. Due to the sparsity of our $B$-band data, \texttt{SNooPy} was unable to determine a $B$-band peak absolute magnitude using a Gaussian Process function, so we instead use the nearest data point as our estimate. This happens to be \~4 days post-peak, so we expect a peak $B$-band magnitude similar to SN\,2012dn if accounting for this. Additionally, correcting for the relatively red color of SN~2021qvo (SNooPy finds a value of $E(B-V)_{host} =0.151\pm0.037$~mag) would increase its absolute magnitude by $\sim$0.5--1~mag in Figure \ref{fig:lwr}.

\subsection{Photometric Sub-Type Identification}

\textcite{Ashall_2020} presented a method to photometrically differentiate between subtypes using the color stretch parameters, $s_{BV}$ or $s_{gr}$,\footnote{The $s_{gr}$ parameter is analogous to $s_{BV}$ but uses $g-r$ instead of $B-V$.} and the difference between the time of $i$-band maximum versus either the $B$-band or $g$-band maximum. 
The time at which the $i$-band peaks relative to other bands is the physical driver behind why this diagnostic works. It traces the speed of the recombination front of the Fe-group elements in the ejecta and, thus, the temperature evolution and adiabatic cooling rate. 
In particular, 1991bg-like, 2002cx-like, and 2003fg-like SNe\,Ia are easily identified in this parameter space because their $i$-band maximum occurs after their $B$-band or $g$-band maximum times. For more extensive discussion of this diagnostic, see \citet{Ashall_2020}.

Figure \ref{fig:ashall_diagram} shows $s_{gr}$ versus the $g$- and $i$-band times of maximum ($t^{i}_{max}-t^{g}_{max}$), with
SN\,2021qvo located in the 2003fg-like SN\,Ia region. Of the 2003fg-like SNe\,Ia from this sample, SN\,2021qvo has the smallest time difference between the $g$ and $i$ bands yet is still consistent with all other 2003fg-likes in the diagnositc, with $i$-band maximum coming $\sim0.7$ days after the $g$-band maximum.

\subsection{Characterizing the Early-Time Bump}
\label{sec:bump_photom}

\begin{figure}[b]
   \centering
   \includegraphics[width=\linewidth]{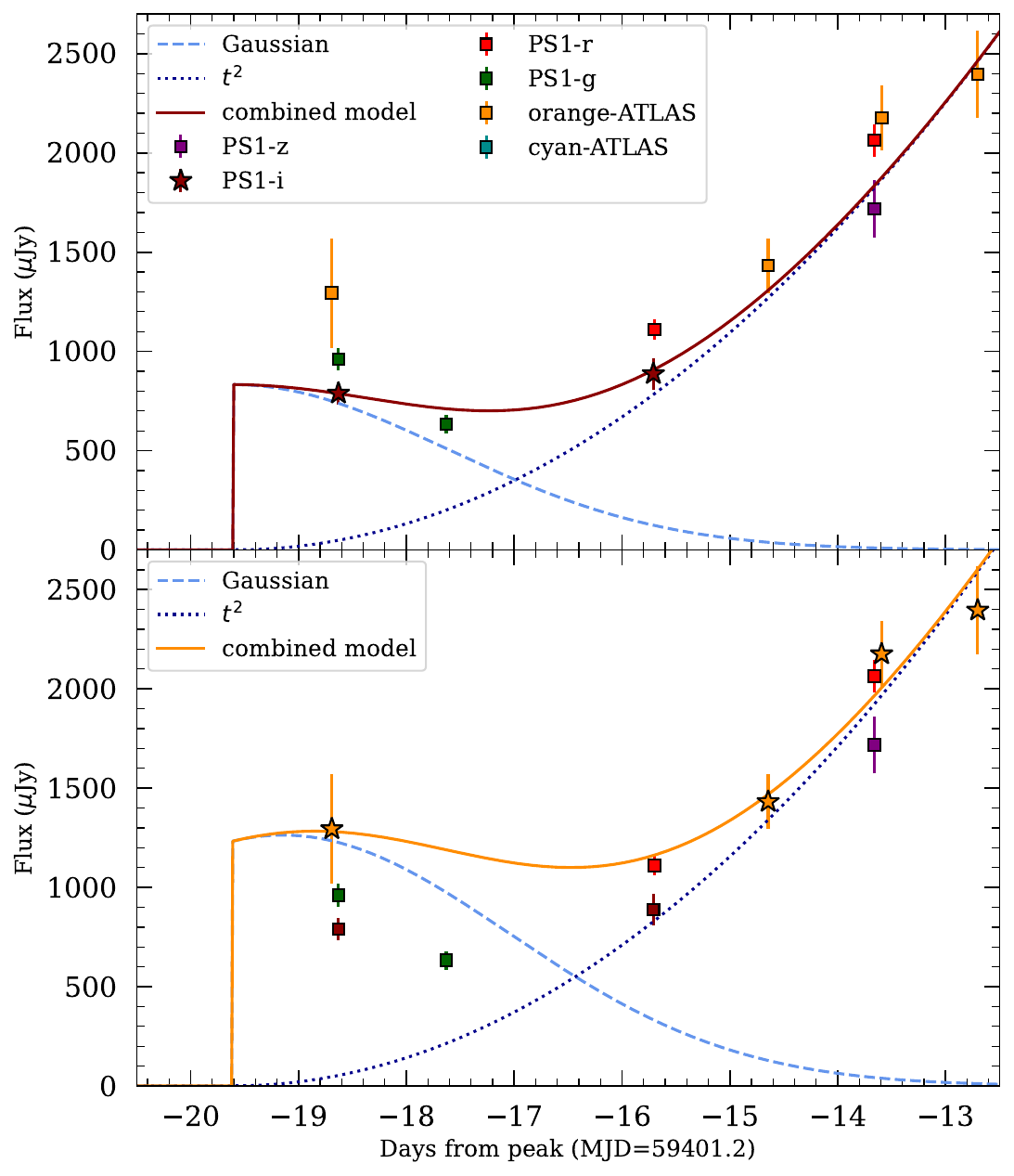}
   \caption{Flux-space light curve of the early time Pan-STARRS and ATLAS data for SN\,2021qvo. For the $i$ (top) and $c$ (bottom) bands, the best-fit Gaussian (dashed) plus exponential $t^2$ rise (dotted) models, using the time of explosion determined by {\tt MOSFiT} (Section \ref{sec:modeling}), are plotted alongside the data. The star symbols indicate the filter fitted with the Gaussian model in each panel.}
   \label{fig:bump_fit}
\end{figure}

As summarized in \citet{Hoogendam_2024}, the rising light curves of SNe\,Ia can be categorized as one of ``single power-law'', ``double power-law'', and ``bump.'' Most SNe\,Ia light curves fall under the ``single'' category, where the light curve behavior is well fit by a single power law. In the case of ``double'' light curves, their best fits can be characterized by broken or two-component power laws. ``Bump'' cases have a non-monotonic rising light curve, with a declining light curve within the timescales of $\sim$1-4 days before rising again to maximum light. These bumps have been observed in the UV and optical for 2003fg-like SNe\,Ia \citep{Jiang21, Dimitriadis23, Srivastav23a} and 2002es-like SNe\,Ia \citep[e.g.,][]{iPTF14atg, Srivastav23b}. 

A bump can be observed for SN\,2021qvo in both the Pan-STARRS and ATLAS data. Plotting the data from both telescopes onto Figure \ref{fig:bump_fit} shows nonmonotonic behavior in the rise of the light curve (the $g$ band has the strongest bump, but lack of data during the exponential rise makes it impossible to fit a single-band Gaussian$+$t$^2$ model similar to previous works (\citealp[e.g.,][]{Dimitriadis19, Shappee_2019, Miller_2020_RisingLCs, Wang24, Ye_2024, Hoogendam_2025a, Hoogendam_2025b}). Beginning with an ATLAS $o$-band observation around 16 days from $i$-band peak, a decrease of nearly 0.8~mag occurs in $g$ band within a day before the quick rise indicative of SN light curves.

\begin{figure}
    \centering
    \includegraphics[width=\linewidth]{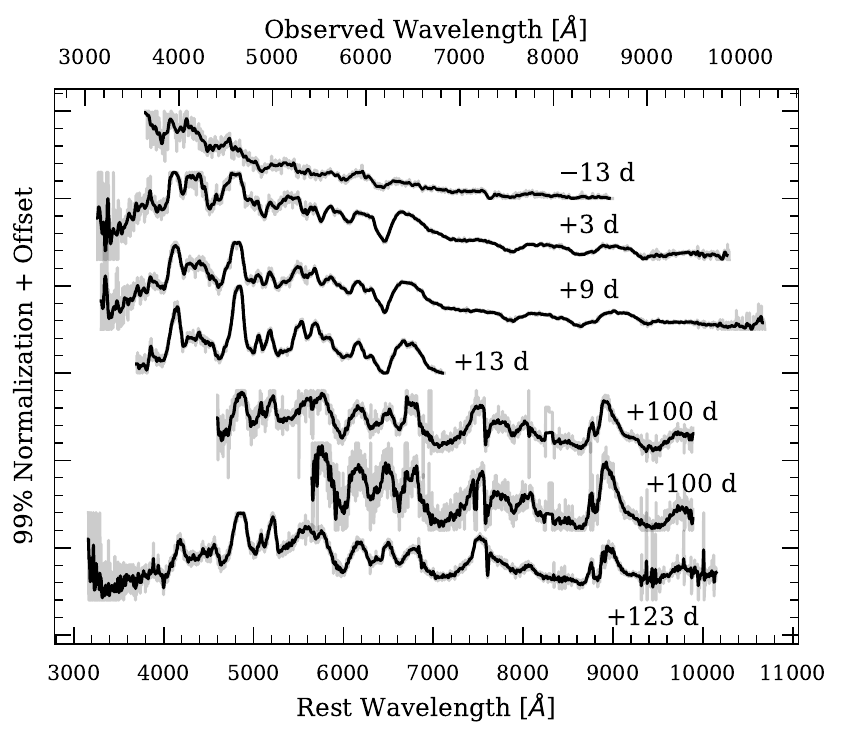}
    \caption{Spectral evolution of SN\,2021qvo. Phases are from the best-fit SNooPy $B$-band peak (MJD 59401.2). Spectra have been normalized by using the range encompassing 99\% of their flux to show features while accounting for the high noise data.}
    \label{fig:spectra_evolution}
\end{figure}

\section{Spectroscopy} \label{sec:spectroscopy}

\begin{figure*}
    \centering
    \includegraphics[width=\linewidth]{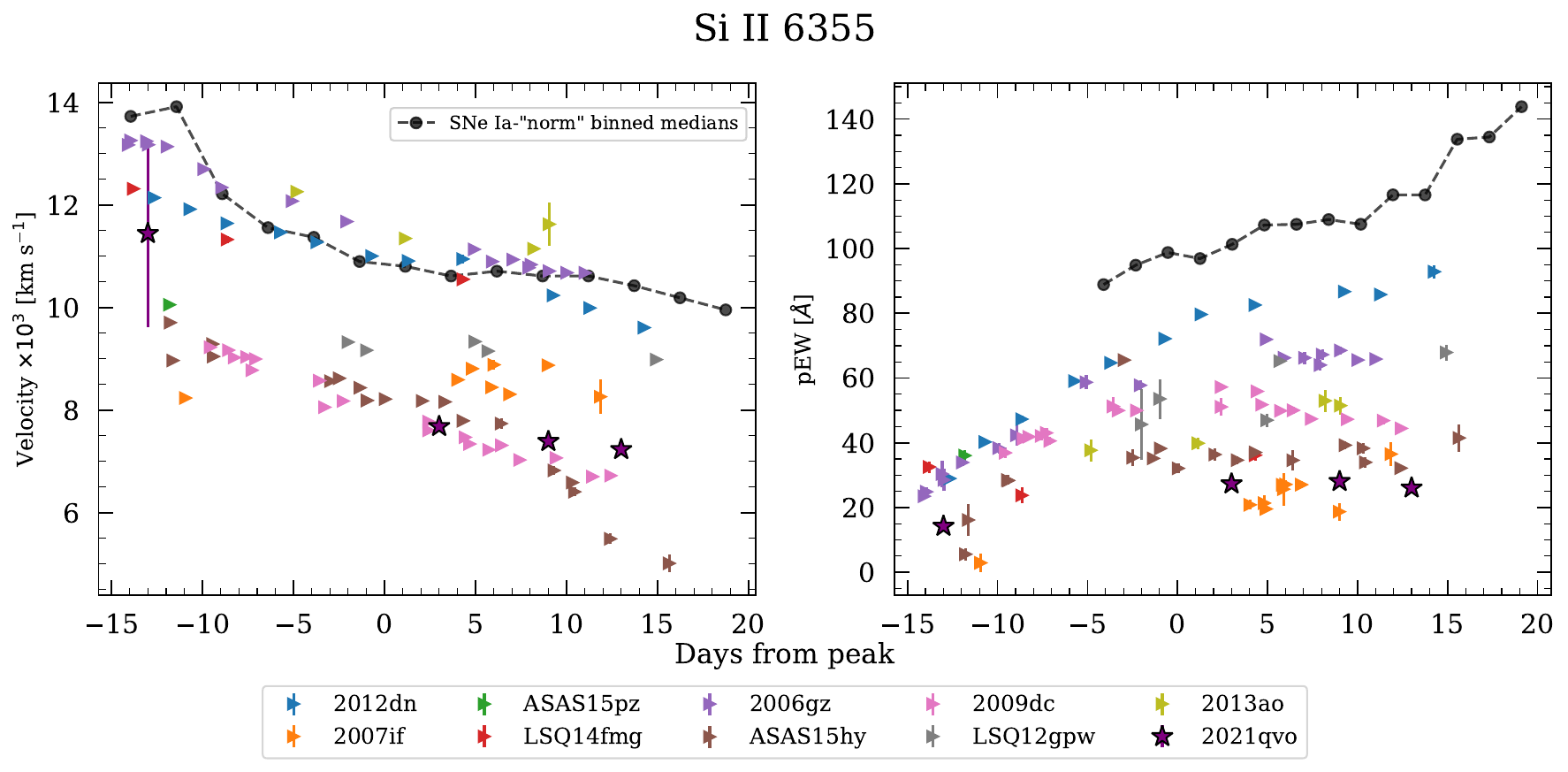}
    \caption{Velocity (left) and pseudo equivalent width (right) of the \SiII\ $\lambda6355$ feature from SN\,2021qvo's spectral data. The pEW and velocity evolutions for other 2003fg-like SNe from the Carnegie Supernova Project catalog \citep{Ashall_2021} are also plotted for comparison. Also plotted in black circles are the binned medians of the normal SNe catalog from \citealp{Kaepora}.}
    \label{fig:vel_pew}
\end{figure*}

The spectroscopic data available for SN\,2021qvo include seven optical spectra taken between $-13$ and $+123$ days from maximum light. These spectra are listed in Table \ref{tab:spec} and their normalized evolution shown in Figure \ref{fig:spectra_evolution}.

\subsection{Spectral Line Measurements}
To quantify the spectral evolution of SN\,2021qvo, we measure velocities and pseudo-equivalent widths (pEWs) for the most prominent spectral features. Both of these quantities were extracted with the semi-automated line fitting \texttt{Python} program, Measure Intricate Spectral Features In Transient Spectra (\texttt{misfits}, \citealp{Holmbo23}\footnote{\href{https://github.com/sholmbo/misfits}{https://github.com/sholmbo/misfits}.}) following the procedure in \citep{Hoogendam_2022}. 

To derive the velocity of a feature, we used the \texttt{velocity.gaussian} function, which determines a pseudo-continuum and then fits a Gaussian profile to the absorption feature, the minimum of which is taken as the velocity. We then use the \texttt{width.shallowpew} function to measure the pEW of a feature with the same pseudo-continuum, which integrates the absorption feature to extract a pEW \citep{garavini_2007}. Both of these functions estimate the respective uncertainties by computing $10000$ realizations using a Monte Carlo approach. The $1\sigma$ standard deviation of the posterior samples serves as the measurement's uncertainty. For both of these measurements, we employed the \texttt{lowpass} and \texttt{rawsmooth} arguments to remove high-frequency noise by passing the spectra through a low-pass filter to smooth them. Figure \ref{fig:vel_pew} shows SN\,2021qvo's \SiII\ spectral velocity and pEW evolutions derived from \texttt{misfits} in comparison to other 2003fg-like SNe and the median values of a normal SNe\,Ia catalog \citep{Kaepora}. The comparative 2003fg-like sample is the same as in Figure \ref{fig:lwr} with the addition of SN\,LSQ12gpw \citep{Baltay_2013a}.

\subsection{The Branch Diagram}

\citet{Branch_2006} found that the maximum light pEWs of the \SiII\ $\lambda 5972$ versus \SiII\ $\lambda 6355$ features differentiate
SN\,Ia into four groups based on their spectral features: core normal, shallow Si, broad line, and cool.  The ``broad line'' SNe~Ia have a broader \SiII\ $\lambda$6355 feature than the ``core normal" SNe~Ia, perhaps indicating a high-velocity component.  ``Shallow silicon'' SNe~Ia have shallower \SiII\ features, especially \SiII\ $\lambda$5972.  Lastly, ``cool'' SNe~Ia have stronger \SiII\ $\lambda$5972 features due to lower temperatures than core-normal SNe\,Ia.

\begin{figure}
    \centering
    \includegraphics[width=\linewidth]{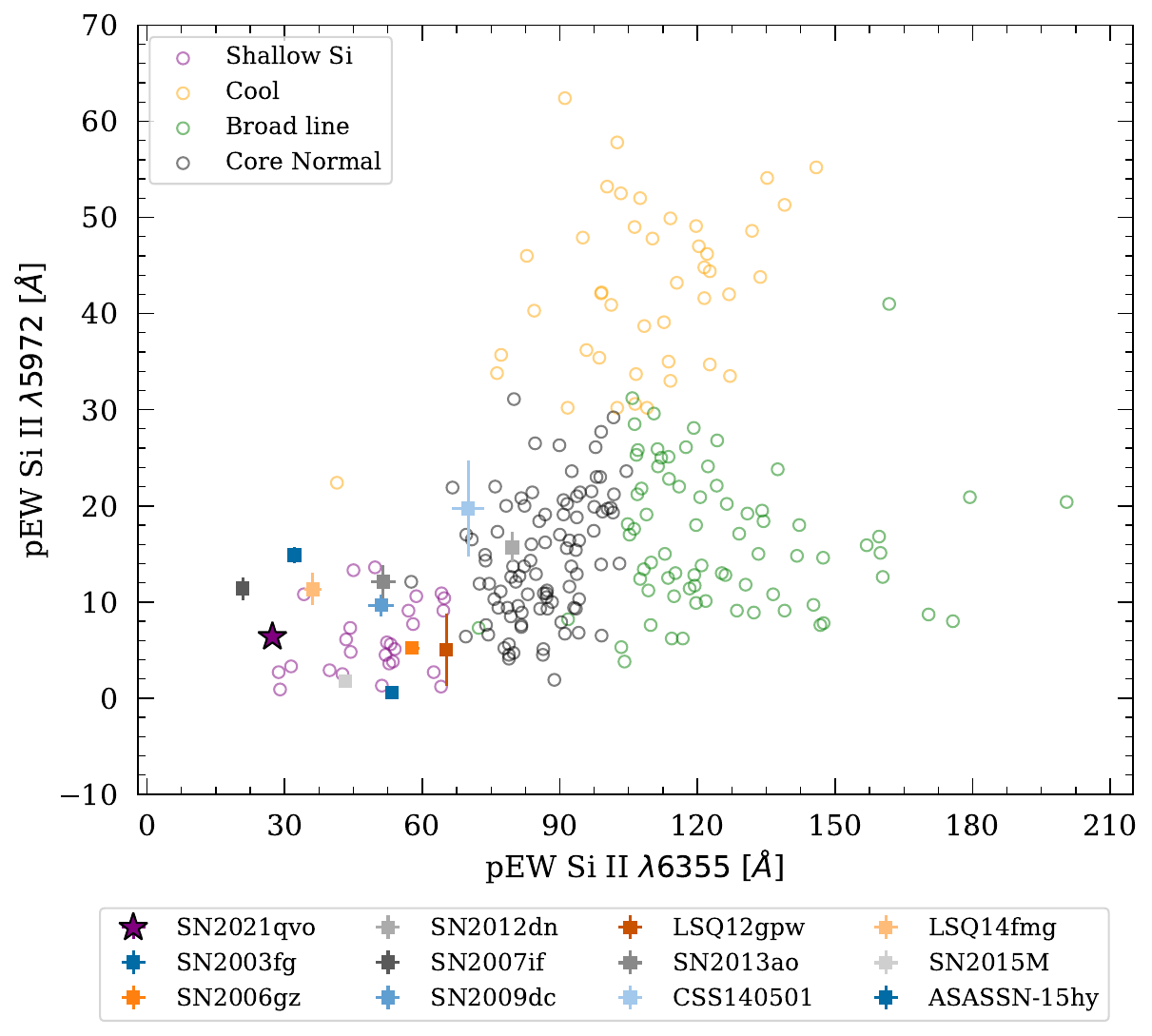}
    \caption{Branch Diagram comparing the pEW of \SiII\ $\lambda5972$ versus \SiII\ $\lambda6355$. The hollow circles represent normal SN\,Ia, with the different colors corresponding to the four groups defined by \textcite{Branch_2006}. Most 2003fg-like SNe\,Ia, including SN\,2021qvo, are in the shallow Si region. Data for the 2003fg-like SNe was taken from \textcite{Ashall_2021}, which includes the SNe samples from Figures \ref{fig:lwr} and \ref{fig:vel_pew} as well as SNe\,2003fg \citep{Howell_03fg}, CSS140501 \citep{Drake_2009}, and 2015M \citep{Morokuma_2015}. The normal SNe\,Ia from \textcite{Blondin_2012}, \textcite{Folatelli_2013}, and \citet{Hoogendam_2022}.}
    \label{fig:branch_diagram}
\end{figure}

Like most 2003fg-like SNe\,Ia \citep{Ashall_2021}, SN\,2021qvo is located in the shallow Si region of the Branch diagram (Figure \ref{fig:branch_diagram}). The \SiII\ $\lambda5972$ pEW feature for SN\,2021qvo is near the average of all 2003fg-like SNe. In contrast, the pEW of the \SiII\ $\lambda6355$ feature is unusually low compared to other 2003fg-like SNe\,Ia events. This could indicate a higher explosion energy for SN\,2021qvo, which may doubly ionize Si and produce \SiIII\ features instead of \SiII. Only the extremely overluminous SN~2007if ($M_V = -20.4$~mag; \citealp{Scalzo10}) has a lower \SiII\ $\lambda6355$ pEW than SN\,2021qvo.

\subsection{Spectral Comparison}

\begin{figure}
    \centering
    \includegraphics[width=\linewidth]{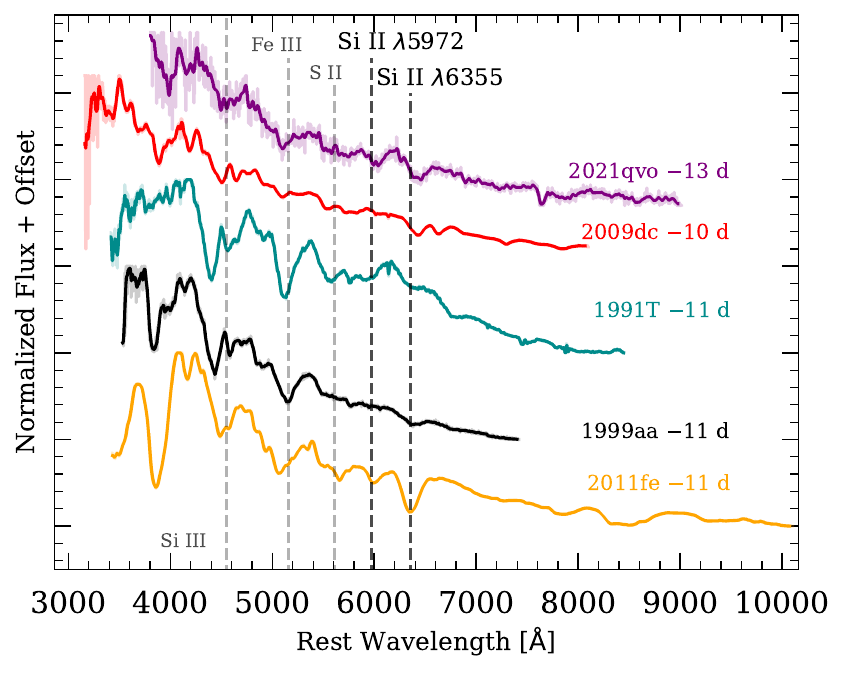}
    \caption{Spectral comparisons to SN\,2021qvo at $-13$ days before maximum light. We show spectra from another 2003fg-like event, SN\,2009dc, as well as a normal SN\,Ia and the notable 1991T and 1999aa subtypes around the same pre-peak epoch. The two notable \SiII\ features are highlighted. Other prominent features in the spectra are also labeled (\SiIII, \FeIII, \SII)}.
    \label{fig:11day_comp}
\end{figure}

\begin{figure}
    \centering
    \includegraphics[width=\linewidth]{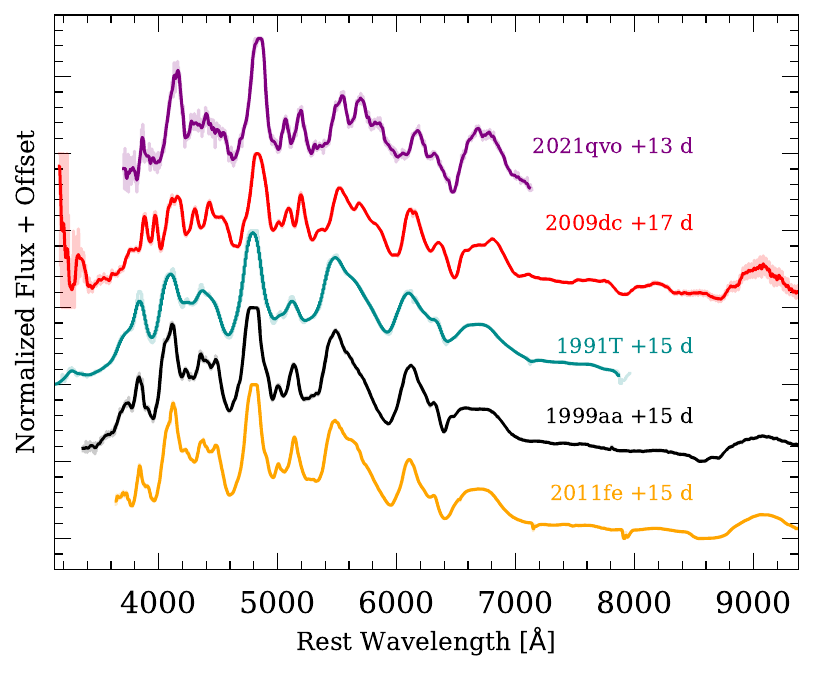}
    \caption{Spectral comparisons to SN\,2021qvo at $+13$ days after maximum light. We show spectra from another 2003fg-like event, SN\,2009dc, as well as a normal SN\,Ia and the notable 1991T and 1999aa subtypes around the same pre-peak epoch.}
    \label{fig:15day_comp}
\end{figure}

We compare SN\,2021qvo to pre- and post-peak spectra from other notable SNe\,Ia in Figures \ref{fig:11day_comp} and \ref{fig:15day_comp}. Figure \ref{fig:11day_comp} compares the $\sim$$-$13~day spectrum of SN\,2021qvo to four other SNe\,Ia with similarly early spectra. Around $-11$~days, the \SiII~$\lambda6355$ feature in the 2003fg-like SNe\,Ia (SNe\,2009dc and SN\,2021qvo) is present, in contrast to other overluminous subtypes such as 1991T and 1999aa-likes where weak \SiII\ features appear after $-$7~days \citep[e.g.,][]{Phillips_2024}. Compared to SN\,2009dc, SN\,2021qvo has a stronger \SiII~$\lambda6355$ feature but lacks the prominent \CII~$\lambda6580$ feature seen in SN\,2009dc. In contrast to SN~2011fe, a canonically normal SN\,Ia with strong \SiII~$\lambda\lambda$5972, 6355 features, the rest of the sample has considerably shallower \SiII double line features, a trait consistent with the pre-peak and near-peak spectral behavior of these overluminous subtypes (See Figure \ref{fig:branch_diagram}).

Figure \ref{fig:15day_comp} compares SN\,2021qvo to the same SNe\,Ia as Figure \ref{fig:11day_comp} at $\sim$13~days post peak. While these SNe\,Ia have spectra that significantly differ at $-13$ to $-10$~days, by $+13$ to $+15$~days they are mostly similar \citep{Ashall_2021}. This is due to the dispersion of the outer layers in these later days, leading to the recession of the photosphere well into the $^{56}\mathrm{Ni}$ region. $^{56}\mathrm{Ni}$ and its products then influence much of the spectra regardless of the subtype, resulting in spectral similarities at these later epochs. 

One difference in the $\sim$15 days post-peak spectra may be in the line profile in the 5200~\AA\ to 5400~\AA\ region. SNe\,2009dc and 2021qvo have two distinct peaks, whereas SNe\,1991T and 1999aa have one peak and a blended decline, and SN\,2011fe has a protruding knee feature. A detailed abundance tomography is beyond the scope of this work, but a population-level comparison of 2003fg-like SNe\,Ia and other normal and overluminous SNe\,Ia to understand the 5200-5400 region may reveal new insights into the explosion symmetry (perhaps these profiles arise from ejecta asymmetries that can be more easily seen in the NIR; \citealp{OHora_2025}), physics (differences in burning may produce different elemental distributions), or temperature (resulting in different ionization fractions).

\section{Constraints and Comparison of CSM in 2003fg-like SNe Ia}\label{sec:discussion}

In this section, we constrain the explosion parameters of SN~2021qvo, including CSM interaction. We then examine our results alongside the CSM parameter constraints of other comparable 2003fg-like SNe in the literature.

\begin{figure*}[t]
    \centering
    \includegraphics[width=\linewidth]{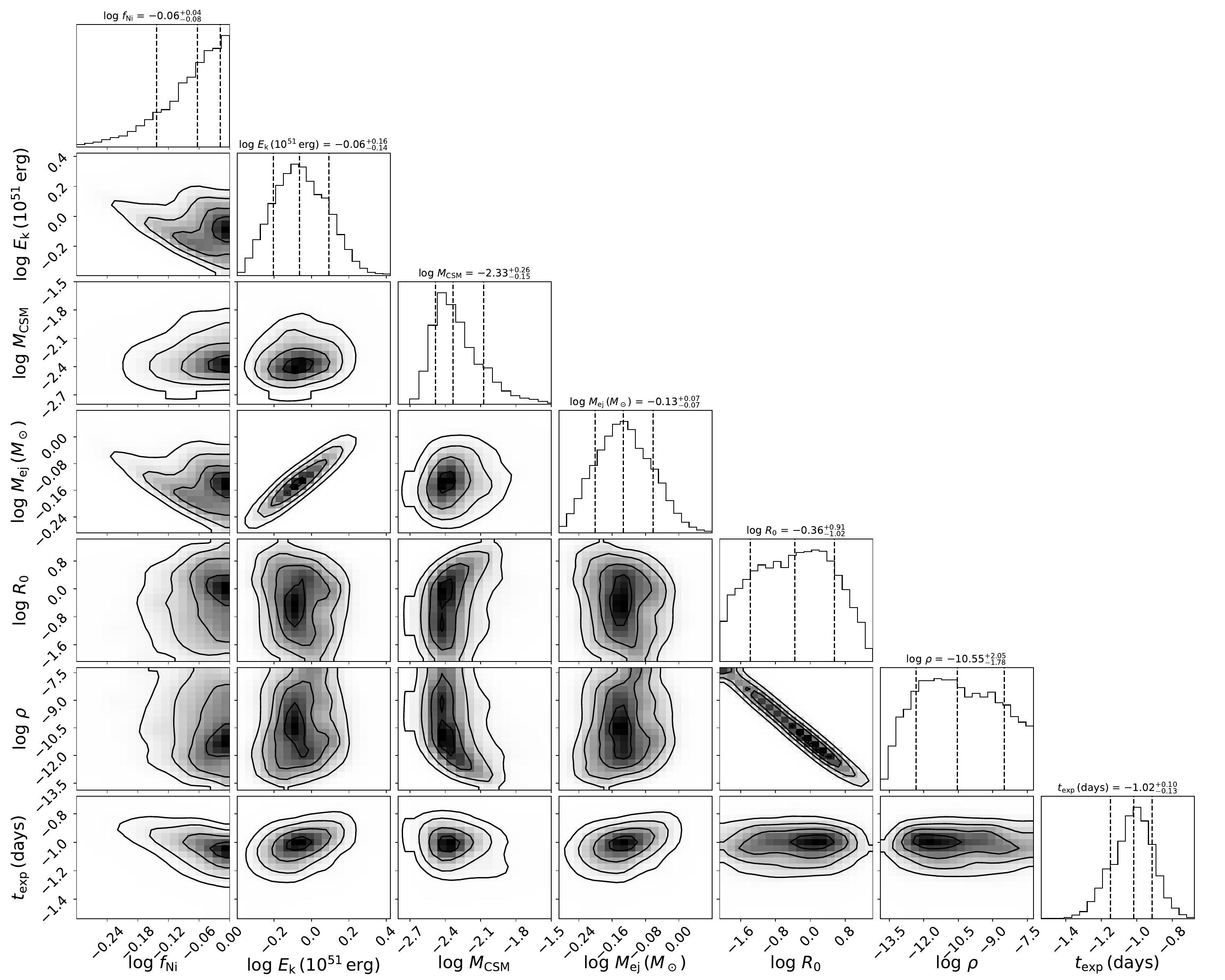}
    \caption{Corner plot of the parameter estimation for the \texttt{MOSFiT} {\tt csmni} model. Parameter constraints (1$\sigma$) are shown above the histogram panels on the right-hand side. 
    See text for explanation of the \texttt{MOSFiT} parameters.}
    \label{fig:MOSFiT_corner}
\end{figure*}

\begin{figure*}
    \centering
    \includegraphics[width=\linewidth]{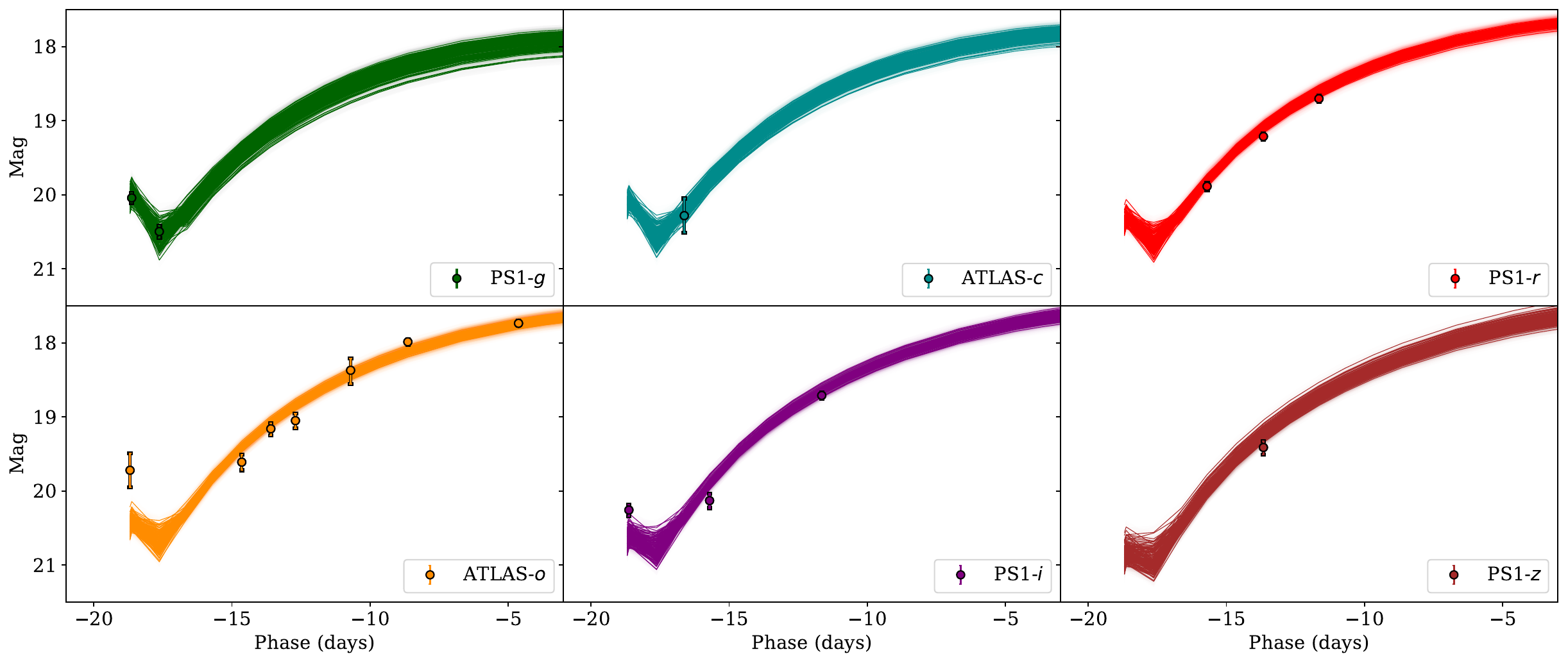}
    \caption{\texttt{MOSFiT} CSM+Ni model fitting on the six photometric bands available for SN\,2021qvo. A bump is observed in the model within the first $\sim$2 days, corresponding to the early-time bump seen in our data. We fit the pre-maximum light photometry up to $-5$~days and use an integrated optical bandpass at later times to constrain the $^{56}$Ni decay (see text).  We note that the post-peak model deviates significantly from the observed colors of SN~2021qvo, as discussed in the text; however, alternate fitting approaches yield consistent CSM parameters.}
    \label{fig:MOSFiT}
\end{figure*}

\subsection{Constraining CSM Interaction} \label{sec:modeling}

We use the Modular Open Source Fitter for Transients (\texttt{MOSFiT}) to model SN\,2021qvo \citep{Mosfit}. \texttt{MOSFiT} is a \texttt{Python}-based package that fits transient data to determine parameter distributions using Monte Carlo methods. Within \texttt{MOSFiT}, we use the built-in \texttt{csmni} model, which uses the CSM-ejecta \citep{Chatzopoulos_2013} and $^{56}$Ni-Co radioactive decay \citep{Nadyozhin_1994} models to fit the light curve of SN\,2021qvo.  In our modeling, we infer parameters by assuming that the early time bump is due to CSM interaction, although see Section \ref{sec:comparisons} for additional discussion.

The \texttt{csmni} model assumes that both ejecta-CSM interaction and $^{56}$Ni decay contribute to the luminosity of the SN\,Ia. A dense CSM may form through mass loss of the progenitor WD \citep{Hachisu_2008}, leftover material from the WD's main sequence stage \citep{Tsebrenko_2015}, or donor material from a companion star \citep{Moriya_2019}. Once the progenitor WD undergoes a SN\,Ia explosion, the SN ejecta interacts with the CSM, resulting in two shock waves, one moving back into the ejecta and the other carrying energy and the CSM outwards. These shocks then ionize the expanding material, resulting in detectable emission and a contribution to the SN luminosity. Such a model may explain overluminous SNe\,Ia by attributing the additional luminosity to CSM interaction \citep{Chevalier_1982, Chevalier_1994, Chatzopoulos_2013}.

For $^{56}$Ni and the subsequent $^{56}$Co decay emission, the primary power source of the SN light curve, the model assumes centrally concentrated $^{56}$Ni. As the WD explodes, the $^{56}$Ni near the core expands outward with the SN ejecta. The $^{56}$Ni then decays to $^{56}$Co, releasing high-energy photons absorbed by the ejecta and re-emitted at lower energies. This primarily powers the observed light curve's peak luminosity and quick rise. The remaining $^{56}$Co then further decays to stable $^{56}$Fe, releasing more photons that power the later phases of the supernova \citep{Chatzopoulos_2012}. This $^{56}$Ni scenario models most normal SN\,Ia well and leads to the  \citet{1982ApJ...253..785A} relation that can be used to estimate $^{56}$Ni mass \citep[e.g.,][]{Stritzinger06}.

At early times, the CSM interaction may produce a significant fraction of the flux; thus, \texttt{MOSFiT}'s \texttt{csmni} combination model can account for early time excess fluxes (see also SN~2022ywc; \citealp{Srivastav23b}). Its constituent CSM interaction and $^{56}$Ni and $^{56}$Co decay models are built from \cite{Chatzopoulos_2013} and \cite{Nadyozhin_1994}, respectively. \texttt{MOSFiT}'s \texttt{csmni} model uses 10 free parameters: total ejecta mass ($M_{\mathrm{ej}}$), the fraction of $^{56}\mathrm{Ni}$ mass with respects to total ejecta mass ($f_{\mathrm{Ni}} = \frac{M_{\mathrm{Ni}}}{M_{\mathrm{ej}}}$), kinetic energy ($E_k$), mass of the CSM shell ($M_{\mathrm{CSM}}$), inner radius of CSM shell ($R_0$), CSM density at inner radius $R_0$ ($\rho_0$), time of explosion relative to first observation ($t_{\mathrm{exp}}$), minimum temperature ($T_{\mathrm{min}}$), host galaxy extinction ($A^{host}_{V}$), and white noise variance ($\sigma$): a variable added to measurement uncertainties to allow additional unmodeled errors \citep{Mosfit}.

We apply a conservative upper bound on $\rho_0$, as it is not well constrained by the data.  We compute the maximum value of $\rho_0$ assuming electron opacity from Thomson scattering and {\tt MOSFiT}'s quadratic power-law density profile.  For a maximum optical depth of $\tau = 5$ --- above which essentially no photons could escape without interaction --- and a Hydrogen CSM, this gives a maximum density of $\rho = 6\times10^{-8}~{\rm g~cm^{-3}}$ when assuming a minimum CSM radius of 2000~km (approximately equal to the surface of a high-mass white dwarf; \citealp{Caiazzo21}).  

Our other priors follow \citet[their Table 1]{Srivastav23a} with only a few minor differences.  Primarily, $R_0$ and $M_{CSM}$ are slightly more conservative in this analysis; we use a lower limit on $R_0$ of 2000~km and a lower limit on $M_{CSM}$ of $10^{-7}~M_{\odot}$.  Additionally, the lower bound on $T_{min}$ is 2000~K instead of 100~K, $E_{k}$ spans from $0.01\times10^{51}$~erg to $4\times10^{51}$~erg (slightly more conservative), and $M_{ej}$ has a higher maximum of $10M_{\odot}$.  We also use the host-galaxy redshift of $z = 0.04205$ to derive the distance; the luminosity distance assumes a flat $\Lambda$CDM model with $\Omega_m = 0.3$, $\Omega_{\Lambda} = 0.7$, and $H_0 = 70~{\rm km~s~Mpc^{-1}}$.  Although not all parameters are well constrained by our limited data, our conservative choice in priors ensures that uncertainties are propagated accurately to the parameters of interest.

Lastly, because our initial fitting resulted in poor fits to the light curves after peak, we use an integrated light curve to fit the later-time data.  We found that {\tt MOSFiT}'s best fit to the full data set initially used CSM interaction to fit the near-peak light curve, in an attempt to rectify discrepancies of up to $\sim$1~mag in the colors near maximum light, while leaving the early time bump unmodeled.  For this reason, we fit the full light curve up to $-5$ days relative to peak and, for the later-time data, we use {\tt SNooPy} to integrate fluxes with Gaussian process interpolation across the optical bandpasses to produce an integrated optical light curve across (approximately) the $griz$ bands and spanning $\sim$-5 to $+45$~days relative to peak.\footnote{This procedure uses SNooPy's bolometric luminosity computation tool, but here we do not subtract the distance modulus or correct for host-galaxy extinction.  We express the integrated quantities in AB magnitudes.}  The early-time data therefore constrains {\tt MOSFiT}'s CSM model, while the later-time light curve constrains {\tt MOSFiT}'s {\tt default} model, which assumes the average light curve is powered by $^{56}$Ni and $^{56}$Co decay.

Our results from \texttt{MOSFiT} using 20\,000 MCMC steps --- to ensure full sampling of the posterior --- are shown in Figure \ref{fig:MOSFiT_corner}. Although we have significant degeneracies, particularly between the density, the CSM mass, and the radius at which the CSM begins (e.g., a closer-in CSM and higher density produce similar light curves to farther-out CSM and lower density), we find that non-zero CSM mass best explains the observed light curves (Figure \ref{fig:MOSFiT}).\footnote{Note that in SN~2022ywc this degeneracy is less pronounced \citep{Srivastav23b}, and therefore is likely due to our limited data around the time of CSM interaction.}  {\tt MOSFiT} prefers CSM beginning at $\sim$0.05--4~AU, with a best-fit mass of ${\rm log}_{10}(M_{\mathrm{CSM}}/M_{\odot}) = -2.33^{+0.26}_{-0.15}$.  There is also a long, low-probability tail to CSM masses of up to $\sim0.03M_{\odot}$ that is not fully ruled out by our data.  We note that using a less-conservative density prior, such as $\rho = 10^{-11}~{\rm g~cm^{-3}}$ following \citet{Srivastav23a}, changes the mean CSM mass by just 0.09~dex, well within the errors.

We omit the extinction, minimum temperature, and white-noise variance from Figure \ref{fig:MOSFiT_corner} for visual clarity.  The host-galaxy extinction is constrained to $\lesssim$0.4~mag at the 1$\sigma$ level, using priors that range from 0 to 5.6 mag following \citet{Srivastav23b}.  The minimum temperature is ${\rm log_{10}}(T_{\rm min}~K) = 3.85^{+0.05}_{-0.16}$; however, it has a bimodal distribution with peaks at  $\sim3.7$ and $\sim$3.9~dex. Lastly, the variance suggests an added model (or data) uncertainty of $\sim$0.08~mag is needed for a reduced $\chi^2 \simeq 1$.

\subsection{Fitting Variants}
\label{sec:variants}

While the {\tt MOSFiT} fitting results are largely consistent with previous work (see Section \ref{sec:comparisons} below), we derive an unphysical $^{56}$Ni fraction near 1.0; this is likely because the $^{56}$Ni model is not always a good approximation to the colors of 2003fg-like SNe\,Ia.  This high $^{56}$Ni fraction is driven by the near-peak and later-time light curve; excluding the early data for SN~2021qvo does not significantly affect our measurement.  Although fitting the later-time light curves is not the primary goal of this analysis, we perform two additional tests to ensure the robustness of our CSM mass estimates.  
First, we fit {\it only} the pre-peak light curve, finding that the {\tt csmni} model can match the early colors well and returns a consistent (0.9$\sigma$) best-fit of ${\rm log}_{10}(M_{\mathrm{CSM}}/M_{\odot}) = -1.85^{+1.022}_{-0.46}$.

Second, we use \texttt{SnooPy}'s integrated light curve computed from $-5$ to $+45$~days with MOSFiT's {\tt default} model, which assumes the light curve is powered by $^{56}$Ni and $^{56}$Co decay, to derive the $^{56}$Ni fraction.  This yields a constraint on the $^{56}$Ni fraction between 0.53 and 0.65, which we then use as a flat prior in the pre-peak, multi-color light-curve fit. We note that the bolometric fit alone yields a slightly higher ejecta mass of $\sim0.87^{+0.15}_{-0.09}~M_{\odot}$, but it is consistent with the baseline results at $\sim$1$\sigma$ significance. From this analysis, we find ${\rm log}_{10}(M_{\mathrm{CSM}}/M_{\odot}) = -1.66^{+1.04}_{-0.72}$, again consistent with the unconstrained fit and demonstrating that the $^{56}$Ni fraction and CSM mass are not strongly correlated and can be constrained independently (Figure \ref{fig:MOSFiT_corner}). We treat the \texttt{MOSFiT} parameter estimation discussed in Section \ref{sec:modeling} as our baseline result, but note that the early-time results alone may favor a slightly higher CSM mass. 

As a third test, we fix $M_{ej}$ to the Chandrasekhar mass to control for potential model degeneracies in ejecta mass when using the early time light curve.  We find only a minimal change to the CSM mass, with a derived value of ${\rm log}_{10}(M_{\mathrm{CSM}}/M_{\odot}) = -2.26^{+0.42}_{-0.20}$ that differs from our baseline results by only 0.07~dex.  This is further evidence that the derived CSM mass is not highly sensitive to uncertainties in the ejecta mass.

Lastly, including possible {\it HST} detections at late times does not significantly change the fit, with CSM masses on order $10^{-2}$ remaining the best-fit solution. The late-time epochs are therefore consistent with a normally declining Ni$^{56}$-powered light curve.

As an independent check on the high $^{56}$Ni fraction, we also measure the $^{56}$Ni mass and total ejecta mass from the \citet{Khatami19} model following the procedure in \citet{Dimitriadis23}.  We adopt the same parameters as \citet{Dimitriadis23}, including a constant opacity of $\kappa = 0.1~{\rm cm^2~g^{-1}}$ and a value of $\beta = 1.6$ (a dimensionless constant) as suggested by Table 6 of \citet{Khatami19}.  We estimate the peak luminosity of SN~2021qvo with {\tt extrabol} as $2.706 \times 10^{43}~{\rm erg~s^{-1}}$ \citep{Thornton24} and adopt the best-fit time from explosion to maximum light determined by \texttt{MOSFiT} of $-19.6$ days.  

From this model, we derive a $^{56}$Ni mass of $1.78~M_{\odot}$ and an ejecta mass of $1.89~M_{\odot}$ (higher than the best-fit MOSFiT ejecta mass of $0.75~M_{\odot}$), for a $^{56}$Ni fraction of 0.94.  We note that we derive $^{56}$Ni fractions of greater than or nearly one for other 2003fg-like SNe; for SN~2021zny we measure a $^{56}$Ni fraction of 1.8,\footnote{We find a small error in the $^{56}$Ni mass calculation of \citet{Dimitriadis23} that results in a significantly increased value.} for SN~2009dc, we find a $^{56}$Ni fraction of 1.6, and for SN~2012dn we find a $^{56}$Ni fraction of 0.97.  In practice, this may mean that bright and slow-evolving transients like 2003fg-like SNe\,Ia require an additional power source beyond $^{56}$Ni to explain their luminosities.  We leave a more in-depth analysis of this topic for future work.

\subsection{Comparison to Other Rising Light Curve Bump Type Ia Supernovae}
\label{sec:comparisons}

SN\,2021qvo is another SN\,Ia with a non-monotonic, rising light-curve bump and is one of fewer than 10 SNe\,Ia to have such a feature.
It is just the fourth rising light curve bump SN\,Ia with CSM mass estimates.
Previous SNe\,Ia with rising light-curve bumps include iPTF14atg \citep{Cao15, Kromer_2016}, SN~2019yvq \citep{Miller20, Tucker21, Burke21}, SN~2020hvf \citep{Jiang21}, SN~2021zny \citep[][although this claim has been disputed by \citealp{Fausnaugh23}]{Dimitriadis23}, SN~2022ilv \citep{Srivastav23a}, and the 2002es-like SN~2022ywc \citep{Srivastav23b}. We also note that SN~2022pul, although lacking early-time coverage, has nebular-phase data that are consistent with expectations from a CSM-ejecta interaction progenitor model \citep{Siebert_2024}.

\begin{figure}
    \centering
    \includegraphics[width=\linewidth]{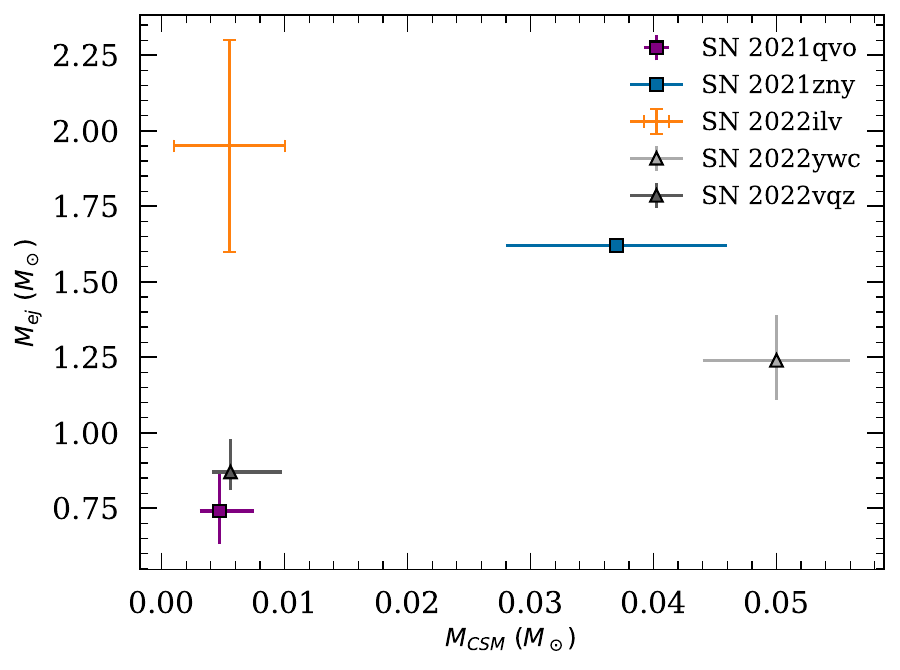}
    \caption{Estimated CSM mass versus SN ejecta mass for early-time bump SNe\,Ia with CSM mass estimates. Here we plot the values for SN\,2021qvo from our \texttt{MOSFiT csmni} results in Section \ref{sec:modeling}. SNe plotted with square markers represent 2003fg-like events, and the triangle marker indicates a 2002es-like SNe. SN\,2022ilv is an 2003fg-like SNe plotted solely with errorbars due to \citet{Srivastav23a} calculating estimated ranges for both the CSM and ejecta masses (\citealp{Srivastav23a} used the \texttt{SuperBol} code from \citealp{Nicholl_2018} to model the bolometric light curves of SN\,2022ilv, then employed the Arnett model to estimate the explosion parameters). Additionally, SN\,2021zny lacks uncertainty in the ejecta mass as \cite{Dimitriadis23} fixed the value for their estimates.}
    \label{fig:csm_ej_comp}
\end{figure}

 The existing sample of early-time bump SNe\,Ia all have CSM mass estimates of order $10^{-2}$ to 10$^{-3}$ $M_\odot$. SN~2022ilv \citep{Srivastav23a} had an estimated $M_\mathrm{CSM} \simeq 10^{-2}$ to 10$^{-3}$ $M_{\odot}$ of material, while SN~2022ywc had $0.050 \pm 0.006 M_{\odot}$ \citep[which had a much stronger and longer-duration bump;][]{Srivastav23b}. SN~2021zny had a best-fit 0.04 $M_{\odot}$ of material \citep{Dimitriadis23}. Our best-fit CSM mass is fully consistent with the mass range from these previous analyses (Figure \ref{fig:csm_ej_comp}).

Of these early-time bump SNe, only SNe~2022ywc and 2022vqz \citep{Xi24}, both 2002es-like SNe, have been fit with the \texttt{MOSFiT} CSM$+$Ni model. 
Overall, while SN~2022ywc has a much stronger shock-cooling signature than 2021qvo, reaching an absolute magnitude of $\sim -19$ in the ATLAS $o$ band, we find that the physical parameters of SN~2021qvo are broadly consistent within the uncertainties. 
Specifically, our ejecta mass, $\sim 0.74 M_{\odot}$, is $\sim$2.7$\sigma$ lower than SN~2022ywc ($\sim1.24 M_{\odot}$), while our explosion energy is consistent within 1$\sigma$.
The CSM density is somewhat higher but just at $\sim$1.2$\sigma$ significance; we emphasize that our measurement from SN~2021qvo is extremely uncertain.
The best-fit CSM radius in SN~2021qvo is also consistent with SN~2022ywc and it is degenerate with CSM mass and density.
SN~2022vqz, which has a relative bump strength that is much more consistent with SN~2021qvo, also has CSM masses that are consistent with SN~2021qvo within the uncertainties ($(M_{\mathrm{CSM}}/M_{\odot}) = 5.6^{+4.2}_{-1.5}\times10^{-3}$) and a slightly higher ejecta mass of $\sim0.87^{+0.11}_{-0.06}~M_{\odot}$.


Previous studies have examined alternative explanations for early-time excesses, including bumps, in SNe\,Ia. These models include surface He-detonations \citep[e.g.,][]{Maeda_2018, Polin_2019}, surface $^{56}$Ni mixing \citep[e.g.,][]{Piro_2016, Magee20a}, and non-degenerate companion interaction \citep[e.g.,][]{Kasen_2010, Maeda_2014}. However, each of these models has challenges to explain the observations, in particular, the blue UV colors and occurrence rates \citep[see][for further discussion]{Hoogendam_2024}. To date, the existing 2003fg-like events with observed light-curve bumps paint a consistent picture of CSM around their progenitor systems, with future statistical samples and fitting methods \citep[e.g.,][]{Sarin24} likely to enable additional tests of this model.

\section{Summary}\label{sec:summary}

In this manuscript, we present SN\,2021qvo, one of a growing class of 2003fg-like SNe\,Ia with bumps in their early-time light curve flux. We analyze photometric data from YSE and ATLAS beginning nearly 17 days before peak, plus an optical spectroscopic time series spanning $-13$ to $+123$~days relative to maximum light. 

SN\,2021qvo occurred in a low-mass dwarf galaxy with ${\rm log(M_{\ast}/M_{\odot}) = 7.83^{+0.17}_{-0.24}}$~dex and has evidence for low stellar metallicity, characteristic of other known 2003fg-like host galaxies. Our analysis of the light-curve shape, luminosity, and \SiII\ pEW and velocities indicate that SN\,2021qvo is fully consistent with the 2003fg-like subtype.

After classifying SN\,2021qvo, we test the leading progenitor theory for overluminous SNe\,Ia with an early time bump, CSM-ejecta interaction. We use the \texttt{MOSFiT} {\tt csmni} model to estimate the interaction of ejecta with CSM in the early time light curve, finding a best-fit CSM mass of ${\rm log}_{10}(M_\mathrm{CSM}/M_{\odot}) = -2.33^{+0.26}_{-0.15}$, which is in broad agreement with constraints from other 2003fg-like SNe\,Ia in the literature. Despite being unable to constrain degeneracies between density and radius fully, we find significant evidence for a non-zero CSM to explain the early-time observed bump.

SN\,2021qvo is the latest example of a 2003fg-like SN\,Ia with a rising light curve bump. The origin of these non-monotonic bumps, which have only been seen in 2002es-like and 2003fg-like SNe\,Ia \citep{Hoogendam_2024}, remains intriguing. Recent observational analyses (e.g., \citealp{Ashall_2021, Hoogendam_2024}) have found the most likely explanation to be SN ejecta interacting with CSM, but the origin of this CSM is not yet well constrained. Possible explanations include an AGB stellar envelope that has lost most if not all of its H and He \citep{Kashi11, Ilkov13, Soker_23}, or from ejected CO during the merger event of two white dwarfs \citep{Shen_2012, Levanon15}. As the number of SNe\,Ia with rising light curve bumps increases due to continued early-time discovery, classification, and follow-up efforts, our understanding of the properties and origin of this plausible enshrouding material will improve.

\section*{Acknowledgments}

We thank Nikhil Sarin for helpful discussions.

I.A.A.P.\ acknowledges support from the Research Experience for Undergraduate program at the Institute for Astronomy, University of Hawai‘i, funded through NSF grant no.\ 2050710, and would like to thank the Institute for Astronomy for their hospitality during the course of this project.
W.B.H.\ acknowledges support from the National Science Foundation Graduate Research Fellowship Program under Grant Nos.\ 1842402 and 2236415. Any opinions, findings, conclusions, or recommendations expressed in this material are those of the author(s) and do not necessarily reflect the views of the National Science Foundation.
D.O.J.\ acknowledges support from NSF grants AST-2407632 and AST-2429450, NASA grant 80NSSC24M0023, and HST/JWST grants HST-GO-17128.028, HST-GO-16269.012, and JWST-GO-05324.031, awarded by the Space Telescope Science Institute (STScI), which is operated by the Association of Universities for Research in Astronomy, Inc., for NASA, under contract NAS5-26555.
The UCSC team is supported in part by NASA grants 80NSSC23K0301 and 80NSSC24K1411; and a fellowship from the David and Lucile Packard Foundation to R.J.F.
C.G.\ is supported by a VILLUM FONDEN Young Investigator Grant (VIL25501) and a Villum Experiment grant (VIL69896). 
M.R.S.\ is supported by a STScI postdoctoral fellowship.

The Young Supernova Experiment (YSE) and its research infrastructure is supported by the European Research Council under the European Union's Horizon 2020 research and innovation programme (ERC Grant Agreement 101002652, PI K.\ Mandel), the Heising-Simons Foundation (2018-0913, PI R.\ Foley; 2018-0911, PI R.\ Margutti), NASA (NNG17PX03C, PI R.\ Foley), NSF (AST--1720756, AST--1815935, PI R.\ Foley; AST--1909796, AST-1944985, PI R.\ Margutti), the David \& Lucille Packard Foundation (PI R.\ Foley), VILLUM FONDEN (project 16599, PI J.\ Hjorth), and the Center for AstroPhysical Surveys (CAPS) at the National Center for Supercomputing Applications (NCSA) and the University of Illinois Urbana-Champaign.

Pan-STARRS is a project of the Institute for Astronomy of the University of Hawai‘i, and is supported by the NASA SSO Near Earth Observation Program under grants 80NSSC18K0971, NNX14AM74G, NNX12AR65G, NNX13AQ47G, NNX08AR22G, 80NSSC21K1572, and by the State of Hawai‘i. The Pan-STARRS1 Surveys (PS1) and the PS1 public science archive have been made possible through contributions by the Institute for Astronomy, the University of Hawai‘i, the Pan-STARRS Project Office, the Max-Planck Society and its participating institutes, the Max Planck Institute for Astronomy, Heidelberg and the Max Planck Institute for Extraterrestrial Physics, Garching, The Johns Hopkins University, Durham University, the University of Edinburgh, the Queen's University Belfast, the Harvard-Smithsonian Center for Astrophysics, the Las Cumbres Observatory Global Telescope Network Incorporated, the National Central University of Taiwan, STScI, NASA under grant NNX08AR22G issued through the Planetary Science Division of the NASA Science Mission Directorate, NSF grant AST-1238877, the University of Maryland, Eotvos Lorand University (ELTE), the Los Alamos National Laboratory, and the Gordon and Betty Moore Foundation.

Some of the data presented herein were obtained at the W.\ M.\ Keck Observatory, which is a private 501(c)3 non-profit organization operated as a scientific partnership among the California Institute of Technology, the University of California, and the National Aeronautics and Space Administration. The Observatory was made possible by the generous financial support of the W.\ M.\ Keck Foundation.

A major upgrade of the Kast spectrograph on the Shane 3~m telescope at Lick Observatory was made possible through generous gifts from the Heising-Simons Foundation as well as William and Marina Kast. Research at Lick Observatory is partially supported by a generous gift from Google.

The data presented here were obtained in part with ALFOSC, which is provided by the Instituto de Astrofisica de Andalucia (IAA) under a joint agreement with the University of Copenhagen and NOT.

This work has made use of data from the Asteroid Terrestrial-impact Last Alert System (ATLAS) project. The Asteroid Terrestrial-impact Last Alert System (ATLAS) proje  ct is primarily funded to search for near earth asteroids through NASA grants NN12AR55G, 80NSSC18K0284, and 80NSSC18K1575; byproducts of the NEO search include images and catalogs from the survey area. This work was partially funded by Kepler/K2 grant J1944/80NSSC19K0112 and HST GO-15889, and STFC grants ST/T000198/1 and ST/S006109/1. The ATLAS science products have been made possible through the contributions of the University of Hawai‘i Institute for Astronomy, the Queen’s University Belfast, the Space Telescope Science Institute, the South African Astronomical Observatory, and The Millennium Institute of Astrophysics (MAS), Chile.


%

\vspace{5mm}
\facilities{Pan-STARRS. The {\it{HST}} data used in this study can be found at MAST:\dataset[10.17909/6bfs-7x11]{http://dx.doi.org/10.17909/6bfs-7x11}}


\software{astropy \citep{2013A&A...558A..33A,2018AJ....156..123A}, SNooPy \citep{Burns_2011}}




\bibliography{sample631}{}
\bibliographystyle{aasjournal}

\end{document}